\documentclass[twocolumn]{autart}    
\usepackage{enumitem}                           
\usepackage {xcolor}
\usepackage{float}
\usepackage{cuted}
\usepackage{graphicx}          
\usepackage{booktabs}
\usepackage{changes}
\usepackage{cancel}
\usepackage{hhline}           
      \usepackage{cuted}

\usepackage[export]{adjustbox}
\usepackage{tikz}
\usetikzlibrary{shapes,arrows}
\usepackage{amssymb,amsmath,graphicx,multicol,epstopdf}
\usepackage{multirow}
\newtheorem{Example}{Example}[section]
\newtheorem{Remark}{Remark}[section]
\newtheorem{Corollary}{Corollary}[section]
\newtheorem{Definition}{Definition}[section]
\newtheorem{Theorem}{Theorem}[section]
\newtheorem{Proposition}{Proposition}[section]
\newtheorem{proof}{Proof}
\newtheorem{Lemma}{Lemma}[section]
\newtheorem{Assumption}{Assumption}[section]

\newcommand{\BM}{\begin{matrix}}
\newcommand{\EM}{\end{matrix}}

\newcommand{\ba}{\begin{array}}
\newcommand{\ea}{\end{array}}
\newcommand{\be}{\begin{eqnarray}}
\newcommand{\ee}{\end{eqnarray}}
\newcommand{\EQQ}{\begin{eqnarray*}}
\newcommand{\ENN}{\end{eqnarray*}}

\newcommand{\R}{{\mathbb R}}

\def\tt#1#2{\left[\begin{array}{c}#1\\ #2\end{array}\right]}

\newcommand{\cov}[1]{\mathbb{C}ov(#1)}

\newcommand{\bd}{\begin{Definition}
\begin{rm} }
\newcommand{\ed}{ \end{rm}
\end{Definition} }

\newcommand{\bexercise}{\begin{exercise}\vspace{-3mm}
\begin{rm} }
\newcommand{\eexercise}{ \end{rm}
\end{exercise} }

\newcommand{\bappend}{\begin{append}\vspace{-3mm}
\begin{rm} }
\newcommand{\eappend}{ \end{rm}
\end{append} }

\newcommand{\basm}{\begin{Assumption} \begin{rm}}
\newcommand{\easm}{\end{rm} \end{Assumption}}

\newcommand{\bpropty}{\begin{property} \vspace{-3mm}\begin{rm}}
\newcommand{\epropty}{\end{rm} \end{property}}

\newcommand{\bremark}{\begin{Remark}
\begin{rm} }
\newcommand{\eremark}{\hfill$\lozenge$ \end{rm}\end{Remark} }
\newcommand{\bt}{\begin{Theorem} \begin{rm} }
\newcommand{\et}{ \end{rm}
\end{Theorem} }
\newcommand{\bl}{\begin{Lemma} \begin{rm} }
\newcommand{\el}{ \end{rm}
\end{Lemma} }
\newcommand{\bcorollary}{\begin{Corollary} \begin{rm} }
\newcommand{\ecorollary}{ \end{rm}
\end{Corollary} }
\newcommand{\bdefinition}{\begin{Definition}\begin{rm} }
\newcommand{\edefinition}{ \end{rm}
\end{Definition} }
\newcommand{\bproposition}{\begin{proposition} \begin{rm} }
\newcommand{\eproposition}{ \end{rm}
\end{proposition} }
\newcommand{\bexample}{\begin{example} \begin{rm} }
\newcommand{\eexample}{ \end{rm}
\end{example} }

\newcommand{\bproof}{\begin{proof} \begin{rm} }
\newcommand{\eproof}{ \end{rm} \end{proof} }

\begin{document}
\begin{frontmatter}

\title{On Kernel Design for Regularized Non-Causal System Identification\thanksref{footnoteinfo}
 }

\vspace*{-8mm}

    \thanks[footnoteinfo]{A prelimnary version of this paper was presented at the 41st Chinese Control Conference, Heifei, China, July 25-27, 2022. Corresponding author Tianshi Chen.  \\ This work was funded by NSFC under contract No. 62273287, the Shenzhen Science and Technology Innovation Council under contract No. JCYJ20220530143418040, and the Thousand Youth Talents Plan funded by the central government of China.} 
    
    \author[CUHKSZ]{Xiaozhu Fang}\ead{xiaozhufang@link.cuhk.edu.cn},    
    \author[CUHKSZ]{Tianshi Chen}\ead{tschen@cuhk.edu.cn}               
    
    \address[CUHKSZ]{School of Data Science and Shenzhen Research Institute of Big Data, The Chinese University of Hong Kong, Shenzhen 518172, China}

\vspace{-4mm}
\begin{abstract} 
Through one decade's development, the kernel-based regularization method (KRM) has become a complement to the classical maximum likelihood/prediction error method and an emerging new system identification paradigm. 
One recent example is its application in the non-causal system identification, and the key issue lies in the design and  analysis of kernels for non-causal systems. In this paper, we develop systematic ways to deal with this issue. In particular, we first introduce the guidelines for kernel design and then extend the system theoretic framework to design the so-called non-causal simulation-induced (NCSI) kernel, and we also study its structural properties, including stability and semiseparability. Finally, we consider some special cases of the NCSI kernel and show their advantage over the existing kernels through numerical simulations. 
\end{abstract}


\end{frontmatter}

\section{Introduction}\label{sec:intro}

In the control community, system identification is the term for the area of building up mathematical models of dynamical systems based on the measured input and output data, and has a history of nearly 60 years \cite{Zadeh:56}. In the first 50 years, the major advance of system identification is on the maximum likelihood/prediction error method (ML/PEM) (and its asymptotic analysis), which is often called the classical system identification, e.g., \cite{Ljung:99,SoderstromS:89,PintelonS:01,Ljung:86}. In the last decade, the major advance is on the kernel-based regularization method (KRM), e.g., the survey papers \cite{PDCDL14,Chiuso16,LCM20} and the book \cite{PCCDL22}. In contrast with ML/PEM, KRM has the feature that it finds a systematic way to engage the prior knowledge of the underlying system to be identified in the system identification loop, in particular in the selection of both the model structure and the model complexity. The carrier of the prior knowledge is the so-called kernel, which determines the model structure, and its parameter (called hyper-parameters) determines the model complexity, which can often be tuned in a continuous way, e.g., \cite{PDCDL14,Chiuso16,LCM20,CHEN18}. Hence, the kernel plays a fundamental role and its design is a key issue. A couple of kernel design methods and many kernels have been proposed, e.g., \cite{PD10,CHEN18,ZC18,MSS16}. Now KRM has become not only a complement to the classical ML/PEM \cite{Ljung:99,COL12}, but also an emerging new system identification paradigm, which is often called the regularized system identification \cite{LCM20,PCCDL22}.



One recent example is the application of KRM in the non-causal system identification \cite{BO20}, where 
the causal tuned-correlated (TC) kernel proposed in \cite{COL12} was first extended to the non-causal case, and then KRM was used to identify the non-causal impulse response and achieved satisfying results. 
The success of KRM in \cite{BO20} is mainly because  KRM can engage, through the designed kernel, the prior knowledge on the non-causal system to be identified, e.g., the stability and smoothness of the non-causal impulse response. This success confirms the efficacy  and also motivates us to further explore the potential of KRM in the non-causal system identification. Then, it is worth noting that all kernels, except the non-causal TC kernel proposed in \cite{BO20}, are causal kernels designed for causal systems, and the design and analysis of non-causal kernels for non-causal systems has not been studied in a systematic way before.

In this paper, we focus on this problem. First, we introduce the  guidelines for the non-causal kernel design and then extend the system theoretic framework proposed in \cite{CHEN18} to design the so-called non-causal simulation-induced (NCSI) kernel. In particular, the NCSI kernel employs the multiplicative uncertainty configuration in robust control, e.g., \cite{ZDG96}, where the nominal model is used to embed the prior knowledge, the uncertainty is modelled by a Gaussian process, and the overall model is simulated with an impulsive input to get the NCSI kernel. Then, we study the structural properties of the NCSI kernel, including stability and semiseparability. Finally, we consider some special parameterizations of the NCSI kernel and show that they give better model estimates/tracking performance than the ones proposed in \cite{BO20} through numerical simulations. 
In contrast with \cite{BO20}, this paper has the following contributions: 
\begin{enumerate}
\item[(i)] we develop a systematic framework to design NCSI kernels for non-causal systems, and moreover, the prior knowledge embedded in NCSI kernels has clear physical interpretation;


\item[(ii)] we study the structural properties of NCSI kernels, including stability and semiseparability;

\item[(iii)] we test the designed NCSI kernels by  Monte Carlo simulations with randomly generated systems and moreover, the simulation results show that the designed NCSI kernels perform (in terms of average accuracy and robustness) significantly better.

\end{enumerate}

The remaining part of this paper is organized as follows. In Section \ref{se:background}, we first introduce some background materials and then the problem statement. In Section \ref{se:guideline}, we introduce the guidelines for non-causal kernel design. In Section \ref{se:ncsi_kernel}, we design the NCSI kernel followed by some analysis and examples. In Section \ref{se:simulation}, we run numerical simulations to illustrate the efficacy of the designed kernels. Finally,  we conclude in Section \ref{se:conclusion}. The proofs of theoretical results are deferred to the appendix.

\section{Background and Problem Statement}\label{se:background}
In this section, we first introduce some background materials and then the problem statement.


\subsection{Non-Causal System Identification}\label{se:non-causal_sysid}
We consider linear time-invariant (LTI), discrete-time, non-causal and bounded-input-bounded-output (BIBO) stable systems described by 
\begin{align}\label{eq:pro_sys1}
y(t)= G(q)u(t)+v(t), t=1, 2, \cdots, N,
\end{align}
where $t$ is the time instant, $u(t),y(t)\in \mathbb{R}$ are the input and output of the system, respectively, $v(t)$ is the measurement noise and assumed to be a white noise with mean 0 and variance $\sigma^2$, $q$ is the forward-shift operator, i.e., $qu(t)=u(t+1)$, and $G(q)$ the transfer function of the system. 
Moreover, $G(q)$ can be represented as
\begin{align}\label{eq:G_0(q) inverse}
G(q)= \sum\limits_{k=-\infty}^{\infty}g^0(k)q^{-k}, 
\end{align} where $\{g^0(k)\}_{k\in\mathbb Z}$ is called the non-causal impulse response of $G(q)$, and clearly, $G(q)$ is BIBO stable if its non-causal impulse response is absolutely summable. 
The non-causal system identification problem is to estimate  $\{ g^0(k)\}_{k\in\mathbb Z}$ or simply $g^0$ below as well as possible based on the input-output data $\{u(t), y(t)\}_{t=1}^N$.

It is worth to stress that the non-causal system \eqref{eq:pro_sys1} naturally arises from the  inverse model control, where the true system is unknown, stable, non-minimum phase and has no zeros on the unit circle, e.g., \cite{BO20} for more details. Besides, the non-causal systems also arise from other contexts, e.g., the continuous-time system identification \cite{GRH21}, and unstable system identification \cite{Fujimoto22}.

\subsection{Kernel-Based Regularization Method}
The non-causal system identification problem of \eqref{eq:pro_sys1} can be handled by the kernel-based regularization method (KRM), which relies on a positive semidefinite kernel $k(t,s; \eta): \mathbb{Z}\times \mathbb{Z}\rightarrow \mathbb{R}$ with $\eta\in\Omega\subset\R^p$ being the hyper-parameter and $p\in\mathbb N$ being its dimension. In particular, KRM searches for a regularized least squares (RLS) estimate $\hat{g}^{\text{R}}$ of $g^0$ 
in the reproducing kernel Hilbert space (RKHS) $\mathcal{H}_k$ associated with $k(t,s; \eta)$: 
\begin{equation}\label{eq:rels_est}
\hat{g}^{\text{R}}= \underset{ g\in \mathcal{H}_k}{\text{argmin}}\sum\limits_{t=1}^N(y(t)-\sum\limits_{k=-\infty}^{\infty}g(k) u(t-k))^2+ \gamma || g||^2_{\mathcal{H}_k},
\end{equation}
where $||\cdot||_{\mathcal{H}_k}$ is the norm of $\mathcal{H}_k$, and $\gamma>0$ is the regularization parameter.

The design of kernels is a core issue for KRM and is referred to as the parameterization of the kernel $k(t,s;\eta)$ through the hyper-parameter $\eta$ by encoding the prior knowledge of the underlying system to be identified.  
The design of causal kernels, i.e., $k(t,s; \eta):\mathbb{N}\times \mathbb{N}\rightarrow \mathbb{R}$, for causal impulse responses has been well studied, e.g., \cite{CHEN18}, and many causal kernels have been introduced, e.g.,  
the diagonal-correlated (DC) kernel and the tuned/correlated (TC) kernel \cite{COL12}:
 \begin{subequations} \label{eq:causal_kernel}
\begin{align}
 &k^{\text{DC}}(t,s;\eta)=c\lambda^{(t+s)/2}\rho^{|t-s|},\label{eq:causal_dc_kernel}\\
&\eta= [c, \lambda, \rho], \quad c\geq 0,\quad  0\leq\lambda<1, \quad |\rho|\leq 1,\nonumber \\
 &k^{\text{TC}}(t,s;\eta)=c\min\{\lambda^{t},\lambda^{s}\}, \label{eq:causal_tc_kernel} \\
 & \eta= [c, \lambda], \quad c\geq 0,\quad  0\leq\lambda<1.\nonumber     
  \end{align}
  \end{subequations} 
In contrast, the design of non-causal kernels, i.e., $k(t,s; \eta):\mathbb{Z}\times \mathbb{Z}\rightarrow \mathbb{R}$, for non-causal impulse responses has not been studied before until recently in \cite{BO20}. Therein, a so-called non-causal TC (NCTC) kernel was introduced and takes the following form
 \begin{align}\label{eq:nctc}
&k^{\text{NC-TC}}(t,s;\eta)= c\min\{b(t), b(s))\}, \\
&\nonumber \qquad b(t) = \left\{ \begin{array}{ll}
\lambda_c^t, &   t\geq 0,\\
\lambda_a^{-t},  & t<0,
\end{array}\right.\\
& \eta= [c, \lambda_c,\lambda_a],\ c\geq 0, \quad 0\leq \lambda_c< 1,0\leq \lambda_a< 1,\nonumber
\end{align} where $c$ is a scale factor, $\lambda_c, \lambda_a$ describe the decay rates for the causal part, i.e., $t\geq 0$ and the anti-causal part, i.e., $t<0$, respectively.

\subsection{Problem Statement}\label{se:pb_for}
As shown in \cite{BO20},  KRM with the proposed kernels, e.g., \eqref{eq:nctc}, can achieve satisfying results for the non-causal system identification, which motivates us to further explore the potential of KRM. In particular, we aim to address the following questions in relation to the kernel design:

\begin{itemize}
\item[1)] what guidelines should be followed when designing non-causal kernels for non-causal impulse responses;
\item[2)] how to design non-causal kernels  in a systematic way  following those guidelines;
\item[3)] is it possible to design non-causal kernels that give even better performance than the ones proposed in \cite{BO20}, e.g., \eqref{eq:nctc}.
\end{itemize}

\section{Non-Causal Kernel Design}\label{se:ncsi_kernel}

The idea is to extend the results for causal kernels to non-causal kernels. 

\subsection{Guidelines for Non-Causal Kernel Design}\label{se:guideline}
Following this idea, we first derive the optimal kernel in the non-causal context and then introduce the guidelines for non-causal kernel design.
\begin{Proposition}[Optimal Kernel, \cite{COL12,PDCDL14}]
Let
$\bar {g}^0$ and $\bar{\hat{g}}$ denote any finite dimensional vector obtained by sampling $g^0(t)$ and its estimate $\hat g(t)$ at the same but arbitrary sampling time instants in  $\mathbb Z$. 
The optimal kernel is defined as 
\begin{align}\label{eq:kernel_opt}
    k^{\text{opt}}(t,s)= g^0(t)g^0(s),\quad t,s\in \mathbb{Z},
\end{align}
which minimizes the MSE matrix 
\begin{align}
MSE(k(t,s))= \mathbb{E}[(\bar{\hat{g}}-\bar {g}^0)(\bar{\hat{g}}-\bar {g}^0)^T],
\end{align} 
in the sense that $MSE(k(t,s))-MSE(k^{\text{opt}}(t,s))$ is positive semidefinite for any positive semidefinite kernel $k(t,s)$, where $\mathbb{E}$ denotes the mathematical expectation.
\end{Proposition}

Similar to the causal case, we have the following guidelines for non-causal kernel design: first, let the kernel mimic the behavior of \eqref{eq:kernel_opt}, and moreover,  the prior knowledge of ${g}^0$ should be used in the kernel design \cite{COL12,CHEN18}; second,
let the kernel have some special structure that can ease the computation of KRM \cite{CALCP14,CA21}. 
For non-causal systems, the most common prior knowledge is the BIBO stability, and besides,
some other common prior knowledge will be introduced later in Section \ref{se:embedmoreprior}.

\subsection{Non-Causal Simulation-Induced Kernel}\label{se:ncsi_kerenl}
Following the guidelines mentioned above, we propose to design kernels for non-causal impulse responses from a system theory perspective. As will be seen shortly, although the same idea has been used in \cite{CHEN18} to design kernels for causal impulse responses, some challenges arise due to the non-causal nature of the problem.


Before proceeding to the details, it is worth sketching the idea of our kernel design method first. Following the first guideline for kernel design, i.e., to mimic the optimal kernel \eqref{eq:kernel_opt}, it is natural to embed the prior knowledge of $G(q)$ from a system theoretic perspective, e.g., properness, stability, dominant dynamics, into a stochastic process $ g(t), t\in \mathbb{Z}$, and then design the kernel as follows 
\begin{align}\label{eq:cov_gtgs}
k(t,s)= \cov{ g(t),g(s)},\ t,s\in \mathbb{Z},
\end{align}
where $\cov{\cdot,\cdot}$ denotes the covariance between two random variables. 
Following this idea, the question now becomes ``how to embed the prior knowledge of $G(q)$ into a stochastic process $ g(t)$"?  

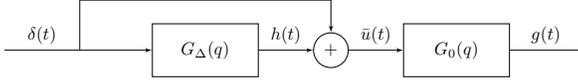
\begin{figure}
   \resizebox{0.45\textwidth}{!}{%
\tikzstyle{block} = [draw, rectangle, 
    minimum height=3em, minimum width=6em]
\tikzstyle{sum} = [draw, circle, node distance=1cm]
\tikzstyle{dio} = [draw, diamond, node distance=1cm]
\tikzstyle{pinstyle} = [pin edge={to-,thin,black}]
\begin{tikzpicture}[auto, node distance=2cm,>=latex']
    \node [coordinate, name=input] {};
    \node [coordinate, right of=input, node distance=1.5cm] (t1) {};      
    \node [coordinate, above of=t1, node distance=1cm] (a1) {};      
    \node [coordinate, right of=a1, node distance=5cm] (a2) {};      

    \node [block, right of=t1, node distance=2.5cm] (t2) {$G_\Delta(q)$};

        \node [sum, right of=t2, node distance=2.5cm] (t3) {$+$};
        \node [block, right of=t3, node distance=2.5cm] (t4) {$G_0(q)$};
    \node [coordinate, right of=t4, node distance=2.5cm] (t5) {};
          \draw [-] (input) -- node {$\delta(t)$} (t1);
          \draw [->] (t1) -- node {} (t2);
          \draw [->] (t2) -- node {$h(t)$} (t3);
          \draw [->]  (t3) -- node {$\bar u(t)$} (t4);
          \draw [-] (t1) -- node {} (a1);
          \draw [-]  (a1) -- node {} (a2);
                    \draw [->]  (a2) -- node {} (t3);
          \draw [-]  (t4) -- node {$ g(t)$} (t5);

\end{tikzpicture}}
\caption{The block diagram for the multiplicative uncertainty. }\label{fig:blk_multi_uncertain}
\end{figure}

One way is to  employ the multiplicative uncertainty framework in robust control, e.g., \cite{ZDG96}, as follows:  
\begin{align}\label{eq:decomp_ft}
G(q)= G_0(q)(1+G_\Delta(q))
\end{align}
where $G_0(q)$ and $G_\Delta(q)$ are called the nominal model and uncertainty, respectively.  For the nominal model $G_0(q)$, we often postulate a classical parameterization $G_0(q,\theta)$ in terms of $\theta$ to embed the prior knowledge of $G(q)$. For example, if the prior knowledge is that $G(q)$ is proper, BIBO stable and its dominant dynamics is over-damped, then we can simply choose $G_0(q,\theta)$ to be a first order system and constrain the values of $\theta$ such that the pole of $G_0(q,\theta)$ is within the unit circle. For the uncertainty $G_\Delta(q)$, we have to suppose as little prior knowledge as possible, and for stable system identification, the only prior knowledge imposed on $G_\Delta(q)$ should be the stability, and in this case, we should design a zero mean Gaussian process to embed such prior knowledge and model the impulse response of $G_\Delta(q)$, then the task of kernel design would be done.  To be more specific, we draw the block diagram of \eqref{eq:decomp_ft} in Fig. \ref{fig:blk_multi_uncertain},  where $\delta(t)$ is a unit impulsive input signal, $\bar u(t),  g(t)$ are the input and output of $G_0(q)$, respectively. Then clearly, if it is possible to construct a zero mean Gaussian process to model the impulse response $h(t)$ of $G_\Delta(q)$ in Fig. \ref{fig:blk_multi_uncertain}, then $\bar u(t)$ and in turn $ g(t)$ would be both Gaussian processes. In this way, we finish the design of the  kernel \eqref{eq:cov_gtgs}, and call \eqref{eq:cov_gtgs} the non-causal simulation-induced (NCSI) kernel.

\subsubsection{The Nominal Model}\label{se:nominal_model}

We will show how to postulate a classical parameterization $G_0(q,\theta)$ of $G_0(q)$ in terms of $\theta$ to embed the prior knowledge of $G(q)$, e.g., properness, stability, dominant dynamics.   
In this section, we will only consider the most fundamental prior knowledge that $G(q)$ is proper and BIBO stable, and later in Section \ref{se:embedmoreprior}, we will consider some more specific prior knowledge.  

As well known from e.g., \cite{Verhaegen96,ZO18}, 
proper and BIBO stable $G_0(q)$ has the following state-space model realization:
\begin{subequations}\label{eq:si_nominal_model}
\begin{align}
&x_c(t+1)= A_cx_c(t)+ B_c\bar u(t),\label{eq:si_nominal_model1}\\
G_0(q, \theta):\quad&x_a(t+1)=  A_a x_a(t)+ B_a\bar u(t),\label{eq:si_nominal_model2}\\
 & g(t)= C_cx_c(t)+ C_ax_a(t)+D\bar u(t),\label{eq:si_nominal_model3}
\end{align}
\end{subequations}
where $A_c, B_c, C_c,  A_a, B_a, C_a, D$ have compatible dimensions, $t\in\mathbb Z$ is the time index, $x_c(t)$ and $x_a(t)$ are the states for the causal part and anti-causal part, respectively, such that all stable poles are contained in $A_c$ and all unstable poles are contained in $A_a$. Let  $\theta=\{A_c, B_c, C_c,  A_a, B_a, C_a, D\}$. Then the state-space model \eqref{eq:si_nominal_model} is a classical parameterization $G_0(q,\theta)$ of $G_0(q)$.

\subsubsection{The Uncertainty}\label{se:uncertainty}


Noting that for stable system identification, the only prior knowledge imposed on $G_\Delta(q)$ should be its stability, we will show how to design a zero mean Gaussian process (or equivalently a kernel) to embed such prior knowledge and model the impulse response $h(t)$ of $G_\Delta(q)$. 
To this goal, since there is no prior knowledge about the smoothness of $h(t)$, it is natural to model $h(t)$ by the following Gaussian process:
\begin{align}\label{eq:un_sibt}
G_\Delta(q): \quad h(t)= b(t)w(t), \quad  t\in \mathbb{Z}, 
\end{align}
where $w(t)$ is a white Gaussian noise with zero mean and unit variance, and $\{b(t)\}_{t\in\mathbb Z}\in\ell_1(\mathbb Z)$. Clearly, $h(t)$ is a zero mean Gaussian process with a non-causal kernel (covariance function) $b(t)b(s)\delta_{t,s}$ with $t,s\in\mathbb Z$, where $\delta_{t,s}$ is the Kronecker delta function, and which is an extension of the causal kernel \cite[eq. (21)]{CHEN18}.

\subsubsection{The NCSI Kernel}\label{se:combined_si_kernel}

Now we put \eqref{eq:si_nominal_model} and \eqref{eq:un_sibt} together and embed them in \eqref{eq:decomp_ft} to get the NCSI kernel \eqref{eq:cov_gtgs}.



\begin{Lemma}\label{le:ss_model} Consider the following state-space model
\begin{subequations}\label{eq:si_ss}
\begin{align}
&x_c(t+1)= A_cx_c(t)+ B_cb(t)w(t),\label{eq:si_expre_causal}\\
&x_a(t+1)=  A_a x_a(t)+ B_a b(t) w(t),\label{eq:si_expre_acausal}\\
& g(t)= C_cx_c(t)+ C_ax_a(t)+  Db(t)w(t),\label{eq:si_expre_causal+acausal}\\
& t= -M_c,-M_c+1,\cdots,\ M_a, \text{for    } M_c,M_a\in\mathbb N,
\end{align}
\end{subequations}
where $-M_c$ and $M_a$ are the starting time for the causal part and anti-causal part, respectively. Assume that the initial states $x_c(-M_c)$ for the causal part and $x_a(M_a)$ for the anti-causal part  are such that $[x_c(-M_c)^T, x_a(M_a)^T]^T$ has zero mean and bounded covariance matrix.  Then for any finite number of sampling time instants $t_1,t_2,\cdots,t_{m}\in \mathbb{Z}$, 
$\lim_{M_c\to\infty}\lim_{M_a\to\infty} [g(t_1),\cdots, g(t_m)]^T$ converges in distribution to a Gaussian random vector with mean zero and covariance matrix, whose $(t,s)$th element is defined through
\begin{subequations}\label{eq:si_expre}
\begin{align}
k^{\text{NCSI}}(t,s)&=\sum\limits_{k=-\infty}^\infty b^2(k) g_0(t-k)  g_0(s-k),\\
& t,s\in \{t_1, t_2, \cdots, t_{m}\},\nonumber
\end{align} 
where $\{g_0(t)\}_{t\in\mathbb Z}$ denotes the non-causal impulse response of the nominal model \eqref{eq:si_nominal_model},i.e.,   
\begin{align}
&g_0(t) =\left\{\begin{array}{ll}
     C_cA_c^{t-1}B_c,& \quad t\geq 1 \\
    D-C_a  A^{-1}_aB_a,    &\quad t=0 \\
       -C_a A_a^{t-1} B_a,    &\quad t\leq -1.
     \end{array}
     \right.,\label{eq:noncausal_g0}
     \end{align}
     \end{subequations}




\end{Lemma}

 \begin{Remark}
 It is worth stressing that although the system theoretic perspective for the design of causal kernels proposed in \cite{CHEN18} is extended for that of non-causal ones
 here, due to the difference in the nature of causal and non-causal systems, the design of non-causal kernel has the combination of the following issues need to be addressed:

 \begin{itemize}
 \item the assumption on $M_c$ and $M_a$: shall they be treated
 as hyper-parameters or chosen to be such that $M_c =
 + \infty$ and $M_a = +\infty$?
 \item the assumption on $x_c(-M_c)$ and $x_a(M_a)$: shall they
 be assumed to be a Gaussian random variable with
 zero mean and a parameterized covariance matrix?
 and shall the mutual or partial independence be assumed on $x_c(-M_c)$, $x_a(M_a)$ and $w(t)$?
 \item the assumption on the uncertainty: the general
 guideline is that the assumption on the uncertainty
 should be made as little as possible (because it is
 uncertainty), but shall the uncertainty be assumed
 to be causal and stable, or anti-causal and stable,
 or non-causal and stable?
 \end{itemize}

 Based on some analysis and moreover, Monte Carlo simulations, we have solutions to the above issues, which are summarized in Lemma 4.1.
 \end{Remark}

\begin{figure}
\centering
\includegraphics[width=0.95\linewidth]{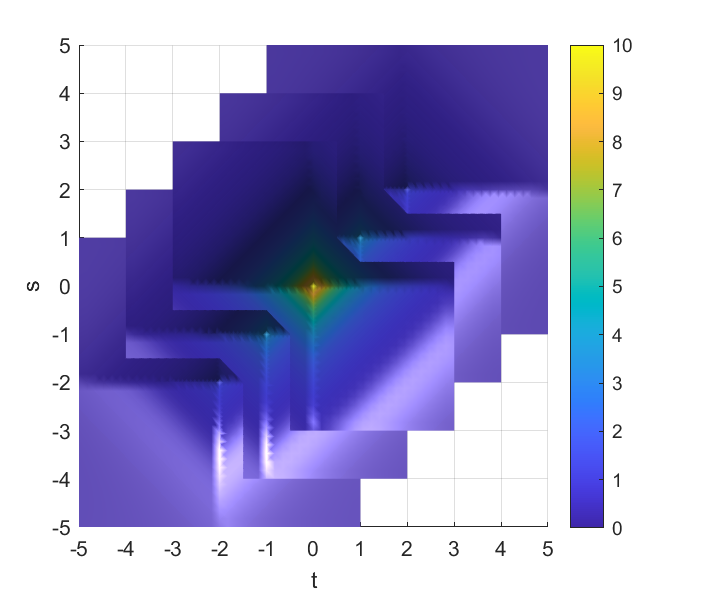}
\caption{Interpretation of the NCSI kernel \eqref{eq:si_expre} as the infinite sum of rank-1 kernels, $g_0(t-k)  g_0(s-k)$, weighted by $b^2(k)$, where we choose $ g_0(t)=10\exp(-|t|)$, and $b(k)=\exp (-|k|)$.  }  \label{fig:stability}
\end{figure}

\begin{Remark} 

Lemma \ref{le:ss_model} shows that as the starting time $M_c$ (actually $-M_c$) for the causal part and $M_a$ for the anti-causal part go to infinity,  then the initial states $x_c(-M_c)$ for the causal part and $x_a(M_a)$ for the anti-causal part will asymptotically have no influence on the NCSI kernel \eqref{eq:si_expre}. This is different from the causal case, where the initial state at starting time $t=0$ has influence on the SI kernel, e.g., \cite[Eq. (22)-(23)]{CHEN18}. 
On the other hand, \eqref{eq:si_expre} shows that the NCSI kernel can be interpreted as an infinite sum of rank-1 kernels, $g_0(t-k)  g_0(s-k)$, weighted by $b^2(k)$, see Fig. \ref{fig:stability}.

\end{Remark}



\subsection{Stability of NCSI Kernel}\label{se:stability}
As can be seen in the following theorem,  the NCSI kernel \eqref{eq:si_expre} is stable in the sense that its associated RKHS $\mathcal{H}_k\subset \ell_1(\mathbb{Z})$, see \cite{CVT06,PDCDL14}, where $\ell_1(\mathbb{Z})$ is  set of absolutely summable sequences of real numbers, indexed by $\mathbb{Z}$.  

\begin{Theorem}\label{th:st}
If  $ \{g_0(t)\}_{t\in\mathbb{Z}}\in \ell_1(\mathbb{Z})$ and $\{b(t)\}_{t\in\mathbb{Z}} \in \ell_1(\mathbb{Z})$, then the NCSI kernel \eqref{eq:si_expre} is stable. 
\end{Theorem}

 \begin{Remark}\label{remark:improvement_stability_si_kernel}
  The technique used in the proof of Theorem \ref{th:st} can be used to prove the stability of the causal SI kernel proposed in \cite{CHEN18}, resulting that the sufficient condition \cite[eq. (29a)]{CHEN18} can be replaced by a weaker one $\{b(t)\}_{t\in\mathbb{N}} \in \ell_1(\mathbb{N})$.
 \end{Remark}

\subsection{Semiseparability of NCSI Kernel}\label{se:semiseparability}
It was shown in \cite{CA21} that, if the kernel for regularized impulse response estimation has some special rank structures, e.g., semiseparability, \cite{VVM08a}, the computational complexity of KRM can be reduced. For example, the widely used DC and TC kernels \eqref{eq:causal_kernel} are both semiseparable, and the simulation-induced kernel for causal impulse response estimation is also semiseparable \cite{CA21}. Interestingly, we will show below that, the NCSI kernel \eqref{eq:si_expre} also has some special rank structure and in fact, it is semiseparable plus diagonal.

\begin{Definition}\label{de:semise}
A positive semidefinite kernel $k(t,s):\mathcal{X}\times\mathcal{X}\rightarrow \mathbb{R}$ is said to be extended $p$-semiseparable plus diagonal, if there exist $\mu_i, \nu_i: \mathbb{Z}\rightarrow \mathbb{R}$, $i=1, \cdots, p$, such that 
\begin{align}\label{eq:de_semise}
k(t,s)=\left\{
\begin{array}{lll}
\sum_{i=1}^p \mu_i(t)\nu_i(s),& \quad t> s\\
d(t),& \quad t= s\\
\sum_{i=1}^p \nu_i(t)\mu_i(s),& \quad t< s
\end{array}
\right.
\end{align} 
where $\mathcal{X}$ represents either $\mathbb{N}$ or $\mathbb{Z}$, $p\in \mathbb{N}$, and $d(t)\in \mathbb{R}$ with $t\in \mathcal{X}$. Moreover, if $d(t)$ further satisfies
\begin{align}
d(t)= \sum_{i=1}^p \mu_i(t)\nu_i(t),
\end{align}
then $k(t,s)$ is said to be extended $p$-semiseparable. 
\end{Definition}
Then we show that the NCSI kernel \eqref{eq:si_expre} is extended $p$-semiseparable plus diagonal. 
\begin{Theorem}\label{th:se} Consider the NCSI kernel \eqref{eq:si_expre} and assume that $A_c$ is non-singular. Then 
the NCSI kernel \eqref{eq:si_expre} is extended $\bar p$-semiseparable  plus diagonal with $\bar p \in \mathbb{N}$ and $\bar p \leq p$, where $p$ is the dimension of $A_c$ in  \eqref{eq:si_expre_causal} plus the dimension of $A_a$ in  \eqref{eq:si_expre_acausal}.
 \end{Theorem}
\begin{Corollary}\label{co:se}  Consider the NCSI kernel \eqref{eq:si_expre} and assume that $A_c$ is non-singular. Then
the NCSI kernel \eqref{eq:si_expre} is extended $\bar p$-semiseparable, if 
 \begin{align*}
 D=0,\quad \text{ or }\quad D-C_aA_a^{-1}B_a =C_cA_c^{-1}B_c
\end{align*}
 \end{Corollary}
 
  \begin{Remark}
One may wonder what would happen for the case when $A_c$ is singular. In this case, the NCSI kernel \eqref{eq:si_expre} is not semiseparable any more, but its corresponding kernel matrix may still have some rank-structure properties. Since this case is less interesting, the corresponding details will not be given here. 
\end{Remark}

As shown in \cite{MC19}, there are many efficient algorithms on matrix computations for extended $p$-semiseparable plus diagonal kernels. We will illustrate this with a concrete example in the next section.

\subsection{Embedding More Prior Knowledge}\label{se:embedmoreprior}
In this section, we try to embed some more specific prior knowledge of $G(q)$ into the NCSI kernel \eqref{eq:si_expre}.

\subsubsection{Over-Damped Dominant Dynamics}
First, we assume that the  prior knowledge is that the dominant dynamics of $G(q)$ is over-damped  and the function $b(t)$ in the model of the uncertainty $G_\Delta(q)$ in \eqref{eq:un_sibt} is an exponential decay one. Therefore, the state-space model \eqref{eq:si_nominal_model} of $G_0(q)$ is parameterized as follows:
\begin{subequations}\label{eq:kernel_si_9}
\begin{align*}
&A_c= B_c=a_c,\quad A_a^{-1}=  B_a=a_a,\nonumber\\
&C_c=c_c,\quad  -C_aA_a^{-2}=c_a,\quad  D-C_a  A^{-1}_aB_a=c_0,\nonumber\\
&  -1< a_c, a_a< 1,\ -\infty <c_c,c_0, c_a<\infty,
\end{align*}
such that 
\begin{align}\label{eq:ncsi_fo1}
&  g_0(t) =\left\{\begin{array}{ll}
    c_ca_c^{t},& \quad t>0 \\
   c_0,    &\quad t=0 \\
      c_aa_a^{-t},    &\quad t<0
    \end{array}
    \right. ,
\end{align}
and moreover, $b(t)$  is parameterized as follows:
\begin{align}\label{eq:ncsi_fo2}
&b(t)=\left\{
\begin{array}{ll}
\sigma_c\lambda_c^{t/2},&\quad  t>0\\
\sigma_0,&\quad  t=0\\
\sigma_a\lambda_a^{-t/2},&\quad  t<0
\end{array}
\right.,\\
&0< \lambda_c,\lambda_a< 1 ,\  \sigma_c, \sigma_a\geq 0.\nonumber 
\end{align}
As a result, we obtain the following non-causal kernel
\begin{align}
&k^{\text{NCSI-FO}}(t,s;\eta)= \text{\eqref{eq:si_expre}} \text{ with } \eqref{eq:ncsi_fo1} \text{ and } \eqref{eq:ncsi_fo2},\\
&\eta= [a_c, a_a,c_c, c_0,c_a, \lambda_c,\lambda_a,\sigma_c, \sigma_a, \sigma_0].\nonumber 
\end{align}
\end{subequations}
Since the nominal model is a first order system for both the causal and anti-causal part, the kernel \eqref{eq:kernel_si_9} is called the first order NCSI (NCSI-FO) kernel.   The NCSI-FO kernel \eqref{eq:kernel_si_9} has a closed-form expression that can be found in Appendix \ref{ap:ncsifo}. As shown in Remark \ref{re:sigma_0isone}, one of the scaling hyper-parameters $c_c, c_0,c_a, \sigma_c,\sigma_0, \sigma_a$ is redundant and thus, $\sigma_0$ is fixed to be 1 and not treated as a hyper-parameter 
hereafter.

\begin{figure}
\centering
\includegraphics[width=1\linewidth]{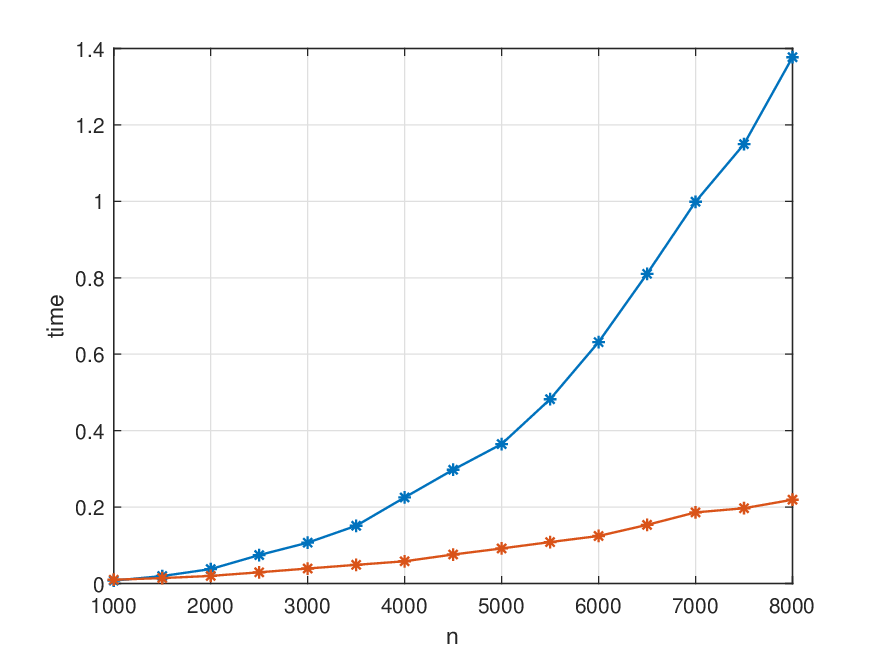}
\caption{The time cost for computing the Cholesky factor of  $K^{\text{NCSI-FO}}$ by using chol in MATLAB (in blue) and by exploiting its high-order semiseparable structure with Algorithm 3 in \cite{MC19} (in red).  }  \label{fig:semisep}
\end{figure}

\begin{Example}
It follows from Theorem \ref{th:se} that, the NCSI-FO kernel \eqref{eq:kernel_si_9} is an extended $2$-semiseparable plus diagonal kernel. Define the kernel matrix $K^{\text{NCSI-FO}}_{ij}= k^{\text{NCSI-FO}}(i,j;\eta_0)$, $ i,j=-n/2, \cdots, n/2$, with $n=1000,\cdots, 8000$. Then Fig. \ref{fig:semisep} shows the time for computing the Cholesky factor of $K^{\text{NCSI-FO}}$ by using chol in MATLAB (in blue) and by exploiting its high-order semiseparable structure with Algorithm 3 in \cite{MC19} (in red). 
 \end{Example}

\subsubsection{Mirrored Poles}
   
Second, we further assume that there is an extra prior knowledge that the dominant dynamics of $G(q)$ has a pair of poles mirrored with respect to the unit circle. As mentioned in \cite[Sec. 4.3]{BO20}, mechanical structures may have such prior knowledge. In this case, it is natural to further let $a_c=a_a$ in the NCSI-FO kernel \eqref{eq:kernel_si_9}, leading to the following kernel
\begin{align}
&k^{\text{NCSI-FO}^\text{mp}}(t,s;\eta)= \eqref{eq:kernel_si_9} \text{ with } a_c=a_a ,\label{eq:kernel_si_9star}\\
&\eta= [a_c, c_c, c_0,c_a, \lambda_c,\lambda_a,\sigma_c, \sigma_a, \sigma_0].\nonumber 
\end{align}

\subsection{Kernel Simplification }\label{sec:simplification}
Since the NCSI-FO kernel \eqref{eq:kernel_si_9} has 9 hyper-parameters, its hyper-parameter estimation may have some difficulties, e.g., its initialization is harder, its local minima issues may be more severe, and it may take more time to find the optimal solution. Hence, following the second guideline for kernel design, i.e., to ease the computation of KRM, we simplify the NCSI-FO kernel \eqref{eq:kernel_si_9} below  by linking some hyper-parameters together.

There are many ways to link the hyper-parameters. 
First, we can link the decay rate of the nominal model with that of the uncertainty as follows:
\begin{align}
&k^{\text{NCSI-DC}}(t,s;\eta)=k^{\text{NCSI-FO}}(t,s;\eta) \text{ with: }\nonumber\\
&\qquad a_c= \rho_c\lambda_c^{1/2},\ a_a= \rho_a\lambda_a^{1/2},\ -1\leq \rho_c, \rho_a\leq  1,\nonumber \\
&\qquad \sigma_c= \sqrt{1-\rho_c^2},\ \sigma_a= \sqrt{1-\rho_a^2},\label{eq:kernel_si_dc}\\
&\eta= [\lambda_c,\lambda_a,c_c, c_0,c_a,\rho_c,\rho_a].\nonumber 
\end{align}
The kernel \eqref{eq:kernel_si_dc} is called the NCSI-DC kernel, because it has a nice property similar to the DC kernel \eqref{eq:causal_dc_kernel}, that is, 
the kernel \eqref{eq:kernel_si_dc} reduces to
 a rank-1 kernel, i.e., $k(t,s)=g_0(t)g_0(s)$, for $\rho_c= \rho_a = 1$, and a diagonal kernel, i.e., $k(t,s)=b(t)b(s)\delta_{t,s}$, for $\rho_c= \rho_a = 0$, which shows different degrees of balance between the nominal model $G_0(q)$ (described by $g_0(t)$) and the uncertainty $G_\Delta(q)$ (described by $b(t)$) by varying $\rho_c$ and $\rho_a$. Moreover, similar to the relation between the TC kernel \eqref{eq:causal_tc_kernel} and DC kernel \eqref{eq:causal_dc_kernel}, we can obtain another special case of the NSCI-DC kernel \eqref{eq:kernel_si_dc} as follows 
\begin{align}
&k^{\text{NCSI-TC}}(t,s;\eta)=k^{\text{NCSI-DC}}(t,s;\eta) \text{ with: }\nonumber\\
&\qquad \rho_c= \lambda_c^{1/2},\ \rho_a= \lambda_a^{1/2},\label{eq:kernel_si_tc}\\
&\eta= [\lambda_c,\lambda_a,c_c, c_0,c_a].\nonumber
\end{align}
which is called the NCSI-TC kernel. 


\begin{Remark}\label{remark:mirrored_poles}
The $\text{NCSI-FO}^\text{mp}$ kernel \eqref{eq:kernel_si_9star} can also be seen as being obtained from the NCSI-FO kernel \eqref{eq:kernel_si_9} by hyper-parameter simplification, i.e., by letting $a_c=a_a$. The difference between the $\text{NCSI-FO}^\text{mp}$ kernel \eqref{eq:kernel_si_9star} and the NSCI-DC kernel \eqref{eq:kernel_si_dc} (also the NSCI-TC kernel \eqref{eq:kernel_si_tc}) is that the extra prior knowledge embedded in the former due to the hyper-parameter simplification $a_c=a_a$ has a clear physical interpretation, i.e., a pair of poles mirrored with respect to the unit circle, but not in the latter.

\end{Remark}

\begin{figure*}
 \centering
 \includegraphics[width=\linewidth]{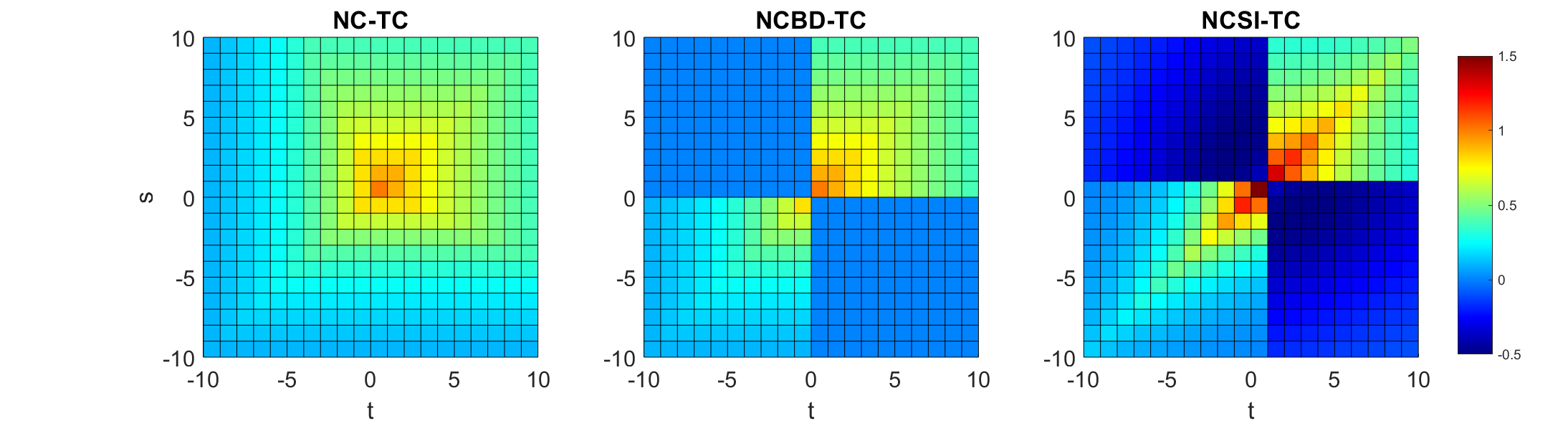}
 \caption{The illustrations of three TC-like non-causal kernels with $\lambda_c= 0.9$, $\lambda_a=0.8$, $c=1$ for the NC-TC kernel \eqref{eq:nctc} and its block diagonal variant, the NCBD-TC kernel \eqref{eq:bdtc}, and $\lambda_c= 0.9$, $\lambda_a=0.8$, $c_c=-1$, $c_0=c_a=1$ for the NCSI-TC kernel \eqref{eq:kernel_si_tc}.  One significant difference of the NCSI-TC kernel \eqref{eq:kernel_si_tc} is its flexibility of choosing negative off-diagonal blocks, which partially accounts for its advantages over the NC-TC kernel \eqref{eq:nctc} and the NCBD-TC kernel \eqref{eq:bdtc}.  }  \label{fig:surf_kernel}
 \end{figure*}

\begin{Remark}\label{re:ncsitc_not_tc}
Despite having the same suffix 'TC', as illustrated in Fig. \ref{fig:surf_kernel}, the NCSI-TC kernel \eqref{eq:kernel_si_tc} has very different behavior from the NC-TC kernel \eqref{eq:nctc} and its block diagonal variant  in \cite{BO20}, 
\begin{align}
     &k^{\text{NCBD-TC}}(t,s;\eta)= \left\{ \begin{array}{ll}
     c\min\{\lambda_c^t, \lambda_c^s\}, &  \text{if } t\geq 0,\ s\geq 0\\
     c\min\{\lambda_a^{-t}, \lambda_a^{-s}\}, &  \text{if } t<0,\ s<0\\
     0, &  \text{otherwise}.
     \end{array}\right.,\label{eq:bdtc}\\
     &\eta= [c, \lambda_c,\lambda_a],\nonumber
\end{align} which is called the NCBD-TC kernel below. Similar to the construction of  the NCBD-TC kernel, we can also construct the NCBD-DC kernel below
 \begin{align}\label{eq:ncbd_dc}
&k^{\text{NCBD-DC}}(t,s;\eta)= 
\left\{ \begin{array}{ll}
c_c^2\lambda_c^{(t+s)/2}\rho_c^{|t-s|}, &  \text{if } t\geq 1, s\geq 1\\
c_0^2, &  \text{if } t= 0, s=0\\
c_a^2\lambda_a^{-(t+s)/2}\rho_a^{|t-s|}, &  \text{if } t\leq -1, s\leq -1\\
0, &  \text{otherwise}
\end{array}\right.,\\
&\eta=[\lambda_c, \lambda_a, c_c, c_0, c_a, \rho_c, \rho_a].\nonumber 
\end{align}

\end{Remark}
 
\section{Numerical Simulation}\label{se:simulation}
In this section, we run simulations to test the kernels proposed in Sections \ref{se:embedmoreprior} and \ref{sec:simplification}, and the ones in \cite{BO20}.



\subsection{Test Systems and Data-Bank}\label{sec:test_systems_and_data}
As mentioned in the last of Section \ref{se:non-causal_sysid}, system \eqref{eq:pro_sys1} often arises from the inverse model control, e.g., \cite{BO20}. More specifically, if the control plant in the feedforward control is denoted by $P(q)$ and assumed to be unknown, stable, non-minimum phase and has no zeros on the unit circle, then its non-causal inverse $G(q)$ can be put in the form of \eqref{eq:pro_sys1}, e.g., \cite[Theorem 1]{BO20}. Moreover, it should be noted that the input $u(t)$ and output $y(t)$ of $G(q)$ correspond to the output and input of $P(q)$, respectively.  

In what follows, we will generate four data banks D1-D4, associated with different types of test systems. We first introduce four types of test systems $P(q)$, then generate the input-output data for each test system $P(q)$, and finally the input-output data for the non-causal inverse of $P(q)$, i.e., $G(q)$ in \eqref{eq:pro_sys1}.

\begin{figure}
\centering
\includegraphics[width=0.95\linewidth]{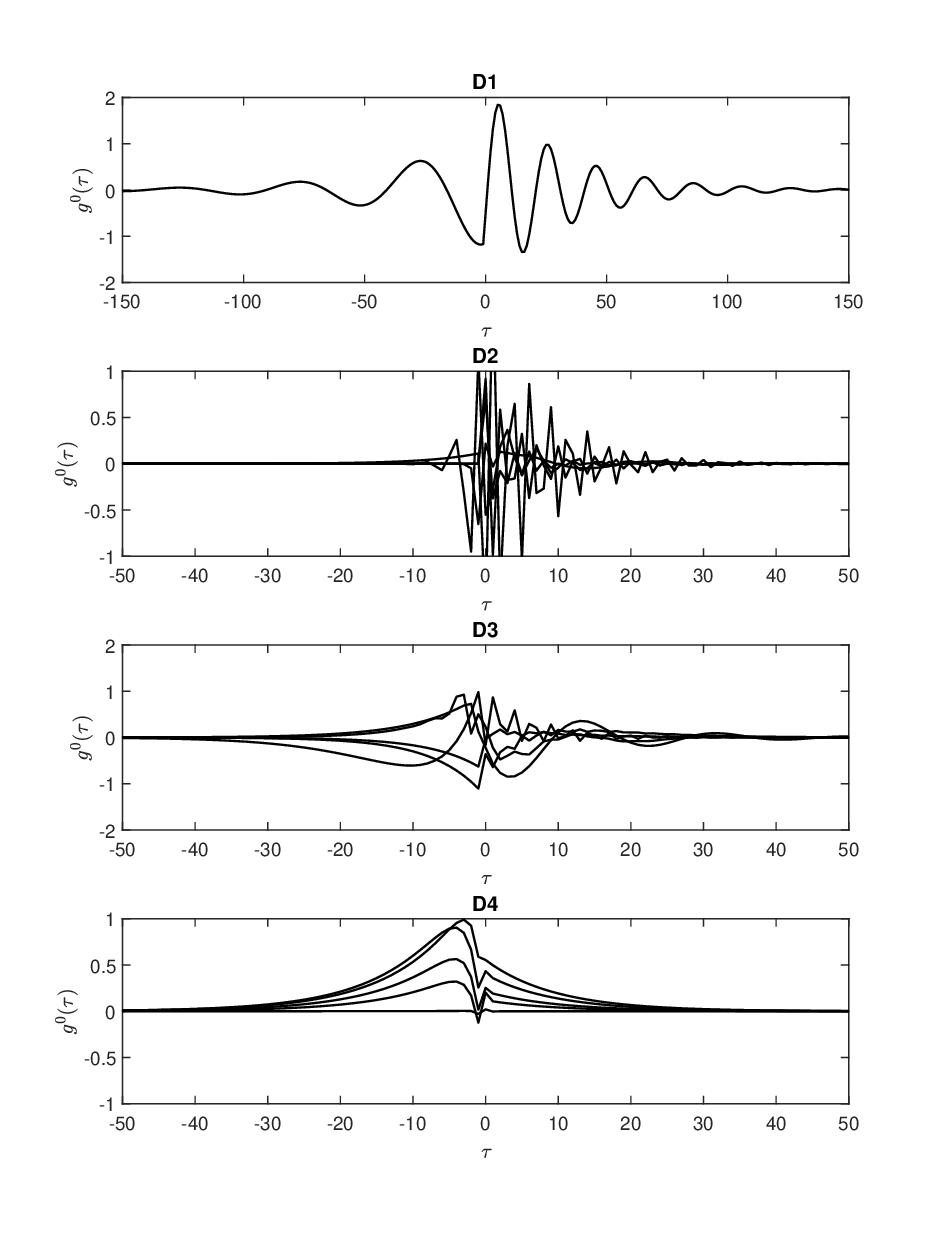} 
\caption{The illustrations of the true non-causal impulse response $ g^0$: the single system \eqref{eq:bosys} in D1, and five system examples in D2-D4.  }  \label{fig:sim2_theta}
\end{figure}

First, we generate the test systems in D1-D4. 
\begin{itemize}
\item Data-bank D1 contains a single non-minimum phase system studied in  \cite{BO20}
\begin{equation}\label{eq:bosys}
\begin{array}{cc}
P(q)= \frac{1.550(q^2-2.035q+1.052)(q^2-1.844q+0.9391)}{q^2(q-0.9514)(q-0.9511)},
\end{array}
\end{equation}
with two stable zeros at $0.922\pm 0.298i$ and two unstable zeros at $1.018\pm 0.126i$. 
\item Data-bank D2 contains 1000 randomly generated systems $P(q)$. We generate each system as follows. First, a 30th order continuous-time system is generated using function \texttt{m=rss(30)} in MATLAB. Then system \texttt{m} is sampled at 100 times of its bandwidth by function \texttt{md=c2d(m,2pi/(100$\cdot$bandwidth(m)))} in MATLAB. Finally, if the system \texttt{md} satisfies the following two conditions:
\begin{enumerate}
\item[(a)] \texttt{md} is a stable, bi-proper and non-minimum phase system, 
\item[(b)] \texttt{md} has all poles and zeros within $\{z\in \mathbb{C}: |z|<0.96 \text{ or } |z|>1.04\}$,
\end{enumerate} 
then \texttt{md} is saved as one of 1000 systems in D2. 

\item Data-bank D3 contains 1000 randomly generated systems $P(q)$.  We generate each system as follows.
For each test system \texttt{md} in D2, we replace the two zeros (can be two real zeros or a pair of complex conjugate zeros) having the smallest magnitude with two new zeros sampled from the uniform distributions, $\mathcal{U}[0.9, 0.8]$ and $\mathcal{U}[1.1, 1.2]$, respectively.
This new system is then saved as one of 1000 test systems in D3 and has the feature that its non-causal inverse $G(q)$ has dominant over-damped dynamics.

Data-bank D4 contains 1000 randomly generated systems. Each test system takes the form of 
\begin{align*}
P(q)= \frac{(q-z_1)(q-1/z_1)(q-z_2)(q-1/z_2)}{(q-p_1)(q-p_2)(q-p_3)(q-p_4)},
\end{align*} 
where $z_1=0.9$, and $z_2, p_1, p_2, p_3, p_4$ are all sampled from the uniform distribution $\mathcal{U}[0, 0.9]$, and thus its non-causal inverse $G(q)$  has mirrored poles.


\end{itemize}
For illustration, some true non-causal impulse responses $g^0(t)$, $t\in\mathbb Z$, of the non-causal inverse $G(q)$ of $P(q)$, are shown in Fig. \ref{fig:sim2_theta}.
The basic prior knowledge that the test systems contain in each data banks of D2-D4 are summarized below follows: 
\begin{itemize}
\item[D2:] BIBO stability and properness;
\item[D3:] BIBO stability, properness, and over-damped dominant dynamics;
\item[D4:] BIBO stability, properness, and over-damped dominant dynamics with mirrored poles.
 \end{itemize}


Second, we generate the input-output data in D1-D4. Each of D1-D4 contains 1000 data sets $\{u(t), y(t)\}^{N_{max}}_{t=1}$, where $N_{max}$ is 2000 for D1 and 700 for D2-D4, respectively. Moreover, we generate 1000 data sets for the single test system \eqref{eq:bosys} in D1, but one data set for each test system in D2-D4. More specifically, the input signal $u(t)$ in \eqref{eq:pro_sys1} is simulated by MATLAB function:
\texttt{u(t)=\text{lsim}(md,$y_0(t)$)} from $t=-n_c$ to $t=N_{max}+n_a$, where $y_0(t)$ is a white Gaussian noise with unit variance. Then the output signal $y(t)$ in \eqref{eq:pro_sys1} is obtained by perturbing $y_0(t)$ with an additive white Gaussian noise $v(t)$ with mean zero and the variance one tenth of the variance of $y_0(t)$. Finally,  $\{y(t),u(t)\}_{t=1}^{N_{max}}$ is saved as one test input-output data for the identification of \eqref{eq:pro_sys1}.




\subsection{Simulation Setup}
We first truncate the non-causal impulse response \eqref{eq:G_0(q) inverse} at sufficiently large orders $n_c$ and $n_a$, and obtain the so-called non-causal finite impulse response (FIR) model
\begin{align}\label{eq:inverse_model_fir}
&y(t)=\sum\limits_{\tau=-n_a}^{n_c}g(\tau) u(t-\tau)+v(t),
\end{align} which can be rewritten into the matrix-vector form as 
\begin{align}\label{eq:matrix_form}
Y= \Psi \theta+ V,
\end{align}
where
\begin{align*}
&Y = [y(n_c+1)\ y(n_c+2)\ \cdots\ y(N-n_a)]^\text{T}, \\
&V = [v(n_c+1)\ v(n_c+2)\ \cdots\ v(N-n_a)]^\text{T}, \\
&\theta = [g(-n_a)\ g(-n_a+1)\ \cdots\  g(n_c)]^T, \\
&\Psi = \left[
\begin{array}{cccc}
u(1+n_c+n_a)&u(n_c+n_a)&\cdots& u(1)\\
u(2+n_c+n_a)&u(1+n_c+n_a)&\cdots& u(2)\\
\vdots & \vdots &\ddots& \vdots\\
u(N)&u(N-1)&\cdots& u(N-n_c-n_a)\\
\end{array}\right]. 
\end{align*}
Accordingly, the regularized least square (RLS) estimate \eqref{eq:rels_est} can be rewritten as  
\begin{align}\label{eq:rls_theta}
&\hat{ \theta}^{\text{R}}= \underset{ \theta\in \mathbb{R}^{n}}{\text{argmin}}||Y-\Psi\theta||_2^2+ \sigma^2\theta^T  K(\eta)^{-1}\theta,
\end{align}
where $\hat\theta^{\text{R}}= [\hat g^{\text{R}}(-n_a)\ \hat  g^{\text{R}}(-n_a+1)\ \cdots\  \hat g^{\text{R}}(n_c)]^T $, $\sigma^2$ is the noise variance of $v(t)$, and $K(\eta)\in \mathbb{R}^{n\times n}$ is a positive semidefinite matrix with its $(i,j)$th entry $K_{i,j}(\eta)$ corresponding to $k(i,j; \eta)$ in \eqref{eq:rels_est}.

The non-causal FIR model orders, $n_c$ and $n_a$, in \eqref{eq:inverse_model_fir} are both chosen to be 150 for D1 and 50 for D2-D4, respectively. 
For each data bank, we will use data sets with the different sample size $N$ with $N\leq N_{max}$, and test the
following kernels:
\begin{itemize}
\item the NC-TC kernel \eqref{eq:nctc}, 
\item the NCBD-TC kernel \eqref{eq:bdtc},
\item the NCBD-DC kernel \eqref{eq:ncbd_dc},
\item the NCBD-TC$^\text{mp}$ kernel, \eqref{eq:bdtc} with $\lambda_c=\lambda_a$,
\item the NCBD-DC$^\text{mp}$ kernel, \eqref{eq:ncbd_dc} with $\lambda_c=\lambda_a$,
\item the NCSI-TC kernel \eqref{eq:kernel_si_tc},
\item the NCSI-DC kernel \eqref{eq:kernel_si_dc},
\item the NCSI-FO kernel \eqref{eq:kernel_si_9},
\item the NCSI-FO$^\text{mp}$ kernel \eqref{eq:kernel_si_9star}.
\end{itemize}
\begin{Remark}\label{re:ncbd-dc}
The NCBD-DC kernel \eqref{eq:ncbd_dc} is proposed by mimicking the structure of the NCBD-TC kernel \eqref{eq:bdtc}. It is worth noting that the NCBD-DC kernel \eqref{eq:ncbd_dc} and the NCSI-DC kernel \eqref{eq:kernel_si_dc} have  the same number of hyper-parameters. 
\end{Remark}

The hyper-parameters are estimated using the empirical Bayes (EB) method, i.e., 
\begin{align}
&\hat{\eta}=  \underset{ \eta\in \Omega_k}{\text{argmin}}\{ Y(\Psi K(\eta) \Psi^T+ \sigma^2 I_{N-n})^{-1}Y+\nonumber\\
&\qquad  \log\det(\Psi K(\eta) \Psi^T+ \sigma^2 I_{N-n})\},\label{eq:eb_opt}
\end{align}
where $\sigma^2$ is estimated from the sample variance of the estimated non-causal FIR model \eqref{eq:inverse_model_fir}  using the LS method. 
The optimization \eqref{eq:eb_opt} is implemented by the MATLAB function \texttt{MultiStart} with \texttt{fmincon} and the number of runs being ten times the number of hyper-parameters.  Substituting \eqref{eq:eb_opt} and the estimate of $\sigma^2$ into \eqref{eq:rls_theta}, we obtain the estimate $\hat\theta^{\text{R}}$. 

The performance of the estimate $\hat\theta^{\text{R}}$ is evaluated by the model fit \cite{Ljung:99} and the tracking error \cite{BO20}. More specifically, the model fit (FIT) is defined by 
\begin{align*}
&\text{FIT} =100\left(1 -
\left[\frac{\sum^{n_c}_{k=-n_a}| g^0(k)-\hat{g}^{\text{R}}(k)|^2
}{\sum^{n_c}_{k=-n_a}| g^0(k)-\bar{g}^0|^2}\right]^{\frac{1}{2}}\right),\nonumber\\
&\bar{g}^0=\frac{1}{n}\sum^{n_c}_{k=-n_a} g^0(k), \nonumber
\end{align*}
where $\{ g^0(k)\}_{k\in\mathbb{Z}}$ is the true non-causal impulse response of $P^{-1}(q)$, and the tracking error (ERR) is defined by
 \begin{align*}
&\text{ERR}= \sqrt{\sum_{k=1}^L e(k)^2/L}, \\
&e(t)=r(t)- P(q)\sum\limits_{k=-n_a}^{n_c} \hat g^{\text{R}}(k)q^{-k}r(t),\\
&t=1-2n_c, \cdots, L+n_a,\nonumber
\end{align*}
where the test length $L=1000$, and the test reference signal $r(t)$ is a white Gaussian noise with unit variance. 
 \begin{figure*}
\centering
\includegraphics[width=0.95\linewidth]{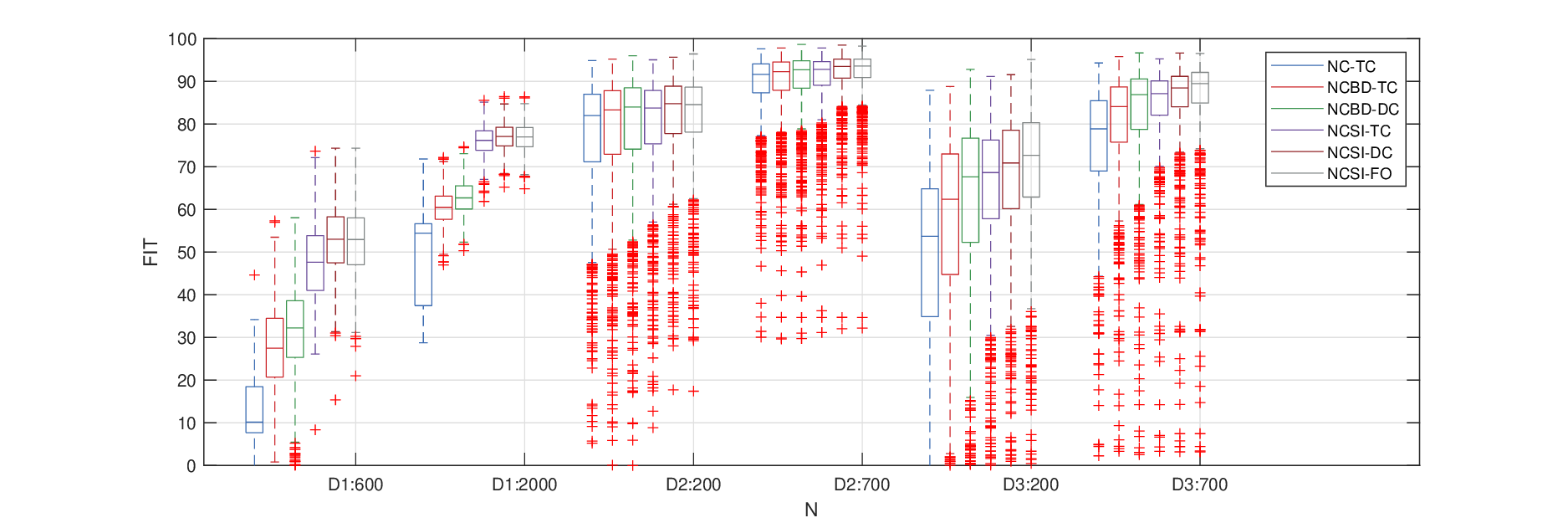}
\includegraphics[width=0.95\linewidth]{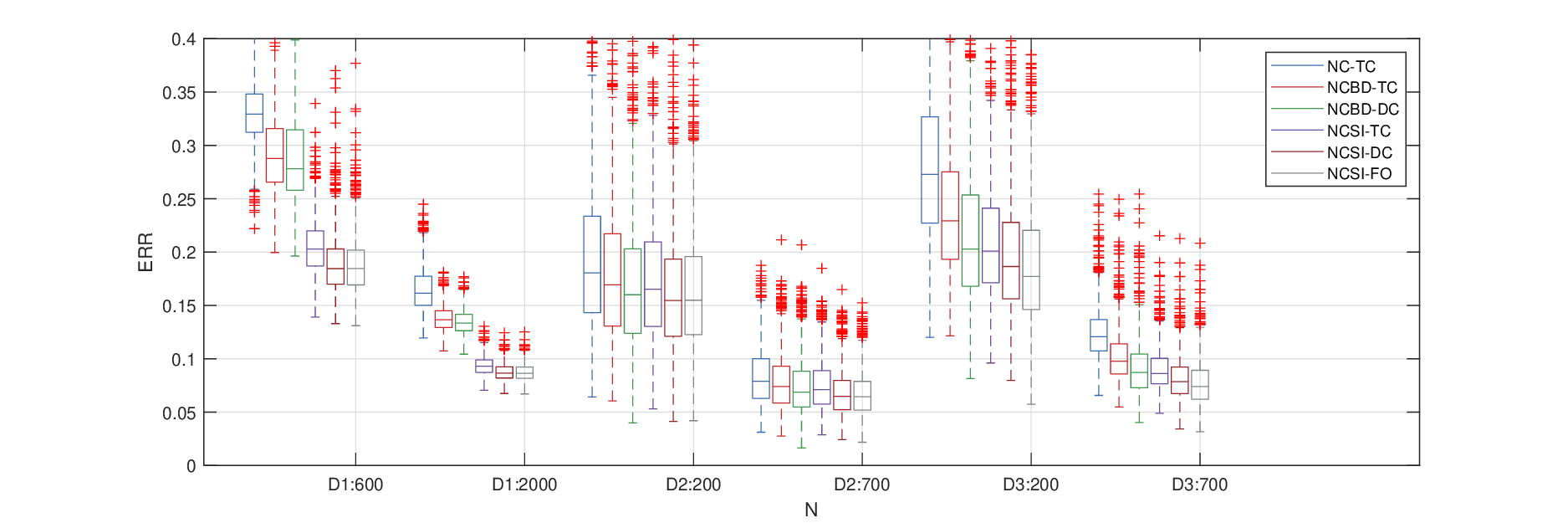}
\caption{Box plots of FITs and ERRs for the data sets in D1-D3.  }  \label{fig:sim1_fit}
\end{figure*}

\begin{figure}
\center
\includegraphics[width=0.95\linewidth]{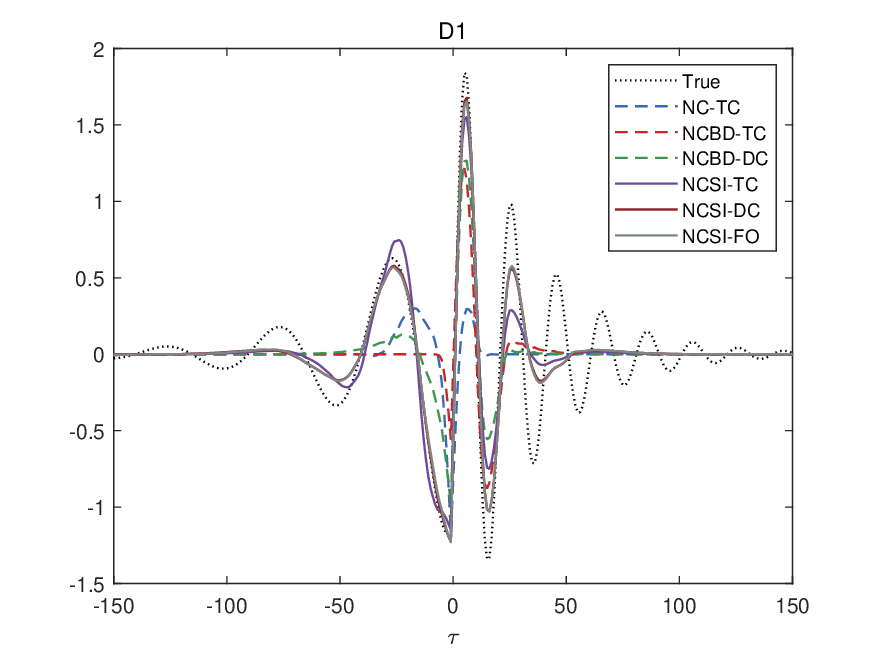}
\caption{Estimates $\hat \theta^{\text{R}}$ for one data set in D1 with $N=600$, and the true non-causal impulse response $\tilde{g}_0$ (dot line). The estimates $\hat \theta^{\text{R}}$ with the NCSI kernels (solid lines) show a better model fit than the existing non-causal kernels (dash lines).   }  \label{fig:sim1_theta}
\end{figure}

\begin{figure}
\includegraphics[width=0.95\linewidth]{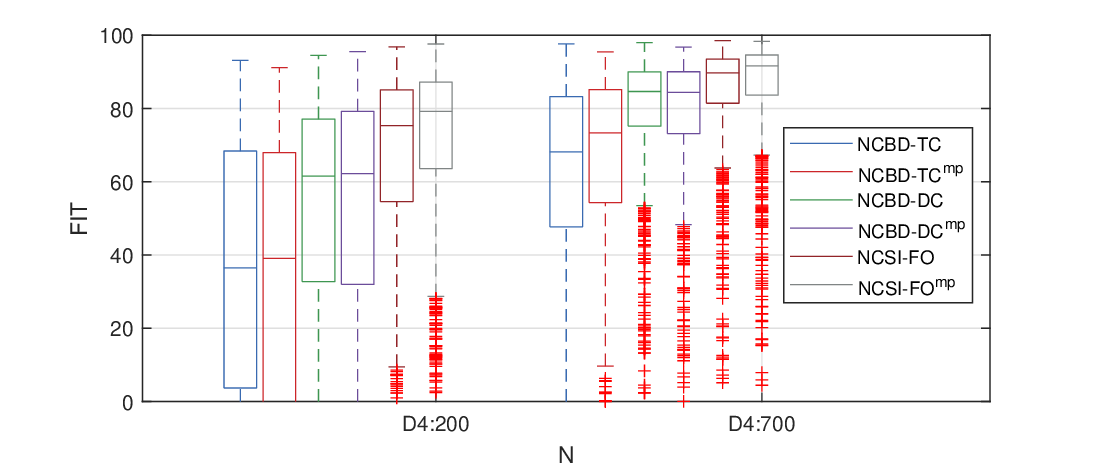}
\includegraphics[width=0.95\linewidth]{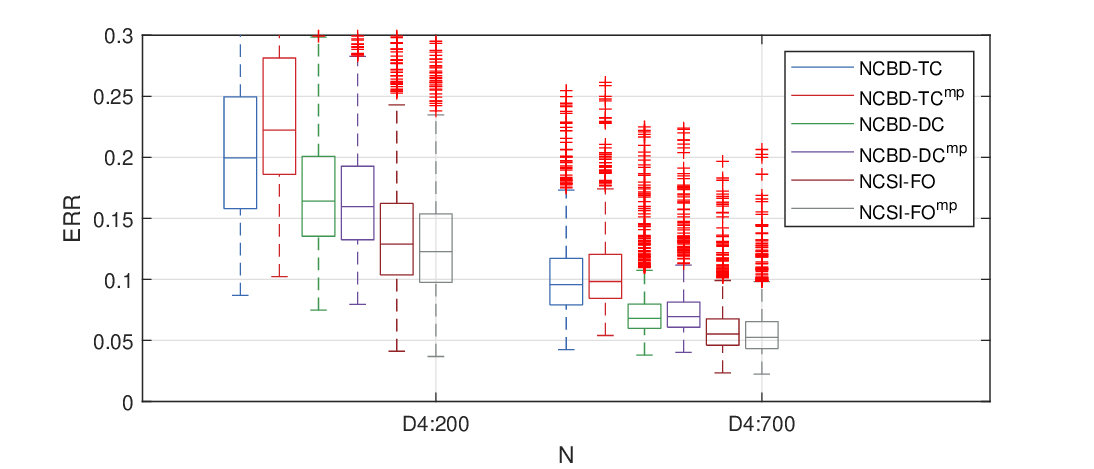}
\caption{Box plots of FITs and ERRs for the data sets in D4.   }  \label{fig:sim4_fit}
\end{figure}

\begin{table*}
  \begin{center}
  \caption{ Average FITs and ERRs, and standard deviation of FITs and ERRs for the data sets in D1-D3.}\label{tab:fiterr_D1}
  \resizebox{ \textwidth}{48mm}{
\begin{tabular}{ccccccccc}
\toprule
\multicolumn{8}{c}{avg. FIT (standard deviation)} \\
\hline
Data-bank&N  & NC-TC&NCBD-TC& NCBD-DC& NCSI-TC  & NCSI-DC& NCSI-FO \\ \hline
D1&600&    12.98(\ \textbf{7.62})        & 27.06(10.47)     &    30.51(11.98)        & 47.64(\ 8.98)    &     \textbf{52.97}(\ {7.77})        & 52.69(\ 8.03)\\
D1&2000&    49.48(10.11)      &   60.41(\ 4.06)     &    62.78(\ 3.80)        & 76.02(\ 3.48)     &    \textbf{76.96}(\ \textbf{3.25})        & 76.85(\ 3.33)\\
D2&200&     76.22(16.61)     &    77.49(16.50)     &   78.62(16.09)        & 78.78(15.15)      &   81.09(\textbf{13.07})       &  \textbf{81.18}(13.14)\\
D2&700&   88.54(10.87)      &   88.98(11.09)         &89.46(11.06)      &   89.87(10.38)       &  91.16(\ 9.65)      &   \textbf{91.28}(\ \textbf{9.61})\\
D3&200&       47.58(23.40)   &      55.39(23.65)      &   59.58(25.26)        & 63.37(\textbf{20.25})        & 65.16(21.29)  &       \textbf{67.20}(21.45)\\
D3&700&   74.76(16.61)  &       79.26(16.40)       &  81.73(16.54)    &     83.13(\textbf{14.93})       &  84.45(15.17)     &    \textbf{85.41}(15.25)\\
\hline
\multicolumn{8}{c}{avg. ERR (standard deviation)} \\
\hline
Data-bank&N  & NC-TC&NCBD-TC& NCBD-DC& NCSI-TC  & NCSI-DC& NCSI-FO \\ \hline
D1&600 & 0.329(0.030)  &  0.291(0.036)    &0.287(0.042)   & 0.205(\textbf{0.025})    &0.189(0.031)  &  \textbf{0.188}(0.029)\\
D1&2000& 0.165(0.020)  &  0.137(0.012)   & 0.134(0.011)    &0.093(0.008)   & 0.087(\textbf{0.007})    &\textbf{0.087}(0.007)\\
D2&200  &  0.194(0.069) &   0.181(0.067)    &0.171(0.067)   & 0.175(0.062)   &\textbf{0.163}(0.058) &0.165(\textbf{0.057})\\
D2&700 &     0.083(0.027)&    0.078(0.027) &   0.073(0.027)  &  0.074(0.024)   & 0.068(0.022)  &  \textbf{0.067}(\textbf{0.022})\\
D3&200 & 0.280(0.073)  &  0.240(0.062)    &0.219(0.072)   & 0.211(\textbf{0.053})    &0.198(0.060)    &\textbf{0.189}(0.063) \\
D3&700&    0.124(0.025) &   0.102(0.024) &   0.091(0.025)  &  0.089(\textbf{0.019}) &   0.081(0.020)  &  \textbf{0.077}(0.021) \\
 \bottomrule
  \end{tabular}}
\end{center}
\end{table*}

\begin{table*}

  \begin{center}
    \caption{Average FITs and ERRs and standard deviation of FITs and ERRs for the data sets in D4.}\label{tab:fiterr_D4}
  \resizebox{\textwidth}{28mm}{
\begin{tabular}{ccccccccc}
\toprule
\multicolumn{8}{c}{avg. FIT (standard deviation)} \\
\midrule
Data-bank&N  &NCBD-TC &NCBD-TC$^{\text{mp}}$&NCBD-DC& NCBD-DC$^{\text{mp}}$& NCSI-FO& NCSI-FO$^{\text{mp}}$\\ \midrule
D4&200&   34.26(37.70)&   32.00(39.83)&   49.17(37.61)&   51.86(33.74)&  64.33(30.08)& \textbf{69.31}(\textbf{27.35}) \\
D4&700&    61.80(27.65)&   64.99(27.23)&   78.15(20.33)&   77.50(20.65)&   83.84(\textbf{16.88})&   \textbf{85.32}(17.13)\\  
\midrule
\multicolumn{8}{c}{avg. ERR (standard deviation)} \\
\midrule
Data-bank&N  &NCBD-TC &NCBD-TC$^{\text{mp}}$&NCBD-DC& NCBD-DC$^{\text{mp}}$& NCSI-FO& NCSI-FO$^{\text{mp}}$\\ \midrule
D4&200&   
    0.208(0.065) &   0.236(0.067)  &  0.178(0.063) &   0.170(0.053)    &0.140(0.053)    &\textbf{0.132}(\textbf{0.049})  \\
D4&700&    0.102(0.034) &   0.106(0.032) &    0.074(0.026)&    0.0761(0.026)&    0.060(\textbf{0.023})&    \textbf{0.058}(0.023)\\   
 \bottomrule
  \end{tabular}}
  \end{center}
\end{table*}

\subsection{Results, Observations and Take-home Messages}\label{se:result_and_finding}
First, we consider the data-banks D1-D3. 
The box plots of FITs and ERRs  are shown in Fig. \ref{fig:sim1_fit}, and the average FITs and ERRs and their standard deviations are shown in Table \ref{tab:fiterr_D1}. For illustration, the estimates $\hat \theta^R$ for one data set in D1 with $N=600$ are shown in Fig. \ref{fig:sim1_theta}. 
Based on these results, we have the following observations:
\begin{itemize}
\item The kernels more to the left in general give better
(and thus the NCSI-FO kernel \eqref{eq:kernel_si_9} in general gives
the best) performance in terms of the average FIT
and ERR, and robustness of the FITs and ERRs
through the distribution and also the standard deviation of the FITs and ERRs. In particular, for
data-banks D1-D2, the NCSI-DC kernel  \eqref{eq:kernel_si_dc} and
NCSI-FO kernel \eqref{eq:kernel_si_9} have very close performance
\end{itemize}
Second, we consider the data-bank D4. The box plots of FITs and ERRs  are shown in Fig. \ref{fig:sim4_fit}, and the average FITs and ERRs and their standard deviations are shown in Table \ref{tab:fiterr_D4}. Then we have the following observations:
\begin{itemize}
\item in contrast with the NCBD-TC kernel \eqref{eq:bdtc} (resp. the NCBD-DC kernel \eqref{eq:ncbd_dc}), the NCBD-TC$^\text{mp}$ kernel (resp. the NCBD-DC$^\text{mp}$ kernel) does not give consistent improvement in the performance in terms of the average FIT and ERR, and robustness of the FITs and ERRs through the distribution and also the standard deviation of the FITs and ERRs; on the other hand, in contrast with the NCSI-FO kernel \eqref{eq:kernel_si_9}, the NCSI-FO$^\text{mp}$ kernel \eqref{eq:kernel_si_9star} gives consistent improvement in the performance in terms of the average FIT and robustness of the FITs and ERRs through the distribution of the FITs and ERRs, but a bit worse standard deviation. 
\end{itemize}
\newpage

Finally, we have the following take-home messages: 
\begin{itemize}
\item in general, the NCSI-FO kernel \eqref{eq:kernel_si_9} is recommended to be used, and if computation issues are
concerned, then the NCSI-DC kernel \eqref{eq:kernel_si_dc} can be
tried as an alternative;
\item if there is extra prior knowledge (e.g., mirrored
poles), then it is suggested to further explore the
structure of the NCSI-FO kernel \eqref{eq:kernel_si_9} and design a
more specific kernel (e.g., the NCSI-FO$^\text{mp}$ kernel)
to embed such prior knowledge
\end{itemize}

\section{Conclusion}\label{se:conclusion}
In this paper, we studied the non-causal system identification problem  by using the kernel-based regularization method. To tackle the key difficulty of non-causal kernel design, we first introduced the guidelines, then extended the system theoretic framework to design the so-called non-causal simulation-induced kernel and studied its properties, including stability and semiseparability. The numerical simulation results showed that the proposed kernels can give better model estimates/tracking performance than the state-of-art kernels. The importance of our obtained results lies in that we develop a systematic framework to design kernels for non-causal systems, the prior knowledge embedded in the designed kernels have clear physical interpretation, and the designed kernels are semiseparable and thus can be used to develop efficient implementations.


\def\thesectiondis{\thesection.}                   
\def\thesubsectiondis{\thesection.\arabic{subsection}.}          
\def\thesubsubsectiondis{\thesubsection.\arabic{subsubsection}.}

\setcounter{subsection}{0}

\renewcommand{\thesection}{A}

\renewcommand{\theequation}{A.\arabic{equation}}
\setcounter{equation}{0}

\renewcommand{\thesubsection}{\thesection.\arabic{subsection}}

\renewcommand{\thesection}{A}
\renewcommand{\thesubsection}{A.\arabic{subsection}}

\section*{Appendix}

\subsection{Proof of Lemma \ref{le:ss_model}}

We will prove Lemma \ref{le:ss_model} in two steps.  First, it is straightforward to show that, given $x_c(-M_c)$, $x_a(M_a)$, $g(t)$ takes the form of
\begin{align}
&g(t)=\left\{
\begin{array}{ll}
&g_{ini}(t)+g_{imp}(t), \text{ if }t,s=-M_c,\cdots, M_a\\
&0, \quad \text{otherwise}
\end{array}\right.,\label{eq:gt_converg_new_variable}\\
&g_{ini}(t)= C_cA_c^{t+M_c} x_c(-M_c)+ C_a A_a^{t-M_a} x_a(M_a),\nonumber\\
&g_{imp}(t)= \sum\limits_{k=-M_c}^{M_a-1}g_0(t-k)b(k)w(k),\nonumber
\end{align}
where $g_0(t)$ is defined in \eqref{eq:noncausal_g0}, and moreover, $\{g_0(t)\}_{t\in \mathbb{Z}}\in\ell_1(\mathbb{Z})$. 
 

Second, we let $Z=[g(t_1),\cdots, g(t_m)]^T$ and then decompose $Z$ as $Z=Z_{imp}+Z_{ini}$, where
\begin{align*}
&Z_{ini}=[g_{ini}(t_1),  \cdots, g_{ini}(t_{m})]^{\text{T}},\\
&Z_{imp}=[g_{imp}(t_1),  \cdots, g_{imp}(t_{m})]^{\text{T}},\\
& -M_c\leq t_1, t_2,\cdots, t_{m}\leq M_a.
\end{align*}
Then we show that as ${M_c\to\infty}$ and ${M_a\to\infty}$,  $Z_{ini}$ converges in probability to a zero vector and $Z_{imp}$
converges in distribution to a Gaussian random vector.


On the one hand, for $Z_{ini}$ and for any $\epsilon>0$, we have 
\begin{align*}
P&\bigg(\|Z_{ini}\|_2>m\epsilon\bigg)\\&\leq P\bigg(|g_{ini}(t_1)|>\epsilon \text{ or }\cdots  \text{ or } |g_{ini}(t_{m})|>\epsilon\bigg)\\
&\leq \sum\limits_{i=1}^{m} P\bigg(|g_{ini}(t_i)|>\epsilon \bigg)\leq \sum\limits_{i=1}^{m}\frac{\cov{g_{ini}(t_i),g_{ini}(t_i)}}{\epsilon^2}
\end{align*} 
where $\|\cdot\|_2$ is the Euclidean norm and the last inequality is due to the Chebyshev's inequality. 
Under the assumption that  $x_{ini}(M_c, M_a)=[x_c(-M_c)^T, x_a(M_a)^T]^T$ has zero mean and bounded covariance matrix $\Sigma_{ini}(M_c, M_a)$,  we have 
\begin{align*}
&\cov{g_{ini}(t_i), g_{ini}(t_i)}= [C_cA_c^{t_i+M_c}\ C_a A_a^{t_i-M_a}] \cdots \\
& \quad \Sigma_{ini}(M_c, M_a)\ [C_cA_c^{t_i+M_c}\ C_a A_a^{t_i-M_a}]^T\rightarrow 0
\end{align*}  
as $M_c, M_a\rightarrow \infty$, and moreover, \begin{align} 
\lim\limits_{M_c\rightarrow \infty }\lim\limits_{M_a\rightarrow \infty }P\bigg(\|Z_{ini}\|_2>m\epsilon\bigg) \leq 0, \label{eq:conv_in_p}
\end{align}
implying that $Z_{ini}$ converges in probability to a zero vector.

On the other hand, for $Z_{imp}$, we consider its characteristic function given by
\begin{align*}
&\varphi(v)= \mathbb{E}[\exp(jv^T Z_{imp})]=\exp(-\frac{1}{2} v^T  \Sigma\ v)\text{ with }\nonumber\\
&( \Sigma)_{i,j} = \cov{g_{imp}(t_i), g_{imp}(t_j)},\ 1\leq i,j\leq m,\nonumber\\
&=\sum\limits_{k=-M_c}^{M_a-1} b^2(k)g_0(t_i-k)g_0(t_j-k).
\end{align*}
Then we have 
\begin{align}
&\lim\limits_{M_c\rightarrow \infty }\lim\limits_{M_a\rightarrow \infty }\varphi( v)=\tilde\varphi( v) =\exp(-\frac{1}{2}  v^T\ \tilde\Sigma\  v)\label{eq:convergence_sigma},\\
&\text{ with }( \tilde \Sigma)_{i,j} =\sum\limits_{k=-\infty}^{\infty} b^2(k)g_0(t_i-k)g_0(t_j-k),\nonumber
\end{align}
where $( \tilde \Sigma)_{i,j}$ exists for all $i,j=1,\cdots, m$ due to the boundedness of $b^2(t)$ and $\{g_0(t)\}_{t\in \mathbb{Z}}\in\ell_1(\mathbb{Z})$. It then follows from the Levy's continuity theorem, e.g., \cite[p. 204]{Loeve78I}, that \eqref{eq:convergence_sigma} leads to  \begin{align}
&Z_{imp}\overset{d}{\longrightarrow} \mathcal{N}(0,\tilde\Sigma),\quad \text{ as } M_c, M_a\rightarrow \infty,\label{eq:conv_in_d}
\end{align}
where $\overset{d}{\longrightarrow}$ denotes the convergence in distribution.

Finally, from the multivariate version of the Slutsky's theorem, e.g., \cite[Thm. 3.4.3]{SS94}, \eqref{eq:conv_in_p} and \eqref{eq:conv_in_d}, it holds that $\lim_{M_c\to\infty}\lim_{M_a\to\infty}Z$ converges in distribution to a Gaussian vector with mean zero and covariance matrix $\tilde\Sigma$. This completes the proof.

\subsection{Proof of Theorem \ref{th:st}}


Note that
\begin{subequations}
\begin{align*}
&\sum\limits_{s=-\infty}^\infty \sum\limits_{t=-\infty}^\infty |k^{\text{NCSI}}(t,s)|\\
&=\sum\limits_{s=-\infty}^\infty \sum\limits_{t=-\infty}^\infty | \sum\limits_{k=-\infty}^\infty b^2(k) g_0(t-k)g_0(s-k)|\\
&\leq \sum\limits_{k=-\infty}^\infty |b^2(k)|\sum\limits_{s=-\infty}^\infty \sum\limits_{t=-\infty}^\infty | g_0(t-k)g_0(s-k)|\\
&= \sum\limits_{k=-\infty}^\infty b^2(k)\sum\limits_{s=-\infty}^\infty \sum\limits_{t=-\infty}^\infty |g_0(t)g_0(s)|\\
&\leq \left(\sum\limits_{k=-\infty}^\infty b^2(k)\right)
 \left(\sum\limits_{k=-\infty}^\infty | g_0(k)|\right)^2<\infty, 
\end{align*}
where the last inequality is true for $ \{g_0(t)\}_{t\in\mathbb{Z}}\in \ell_1(\mathbb{Z})$ and  $\{b(t)\}_{t\in\mathbb{Z}}$ $\in \ell_1(\mathbb{Z})\subset \ell_2(\mathbb{Z})$. Then by \cite[Lemma 3]{DINUZZO15} we complete the proof.

\end{subequations}



\subsection{Proof of Theorem \ref{th:se} and Corollary \ref{co:se}}
Before proceeding to the details, we introduce a lemma.
\begin{Lemma}\label{le:se}
Given three matrices $A\in\mathbb{R}^{n\times n}$, $B\in\mathbb{R}^{n\times 1}$ and $C\in\mathbb{R}^{1\times n}$ with $A$ being non-singular, there exist two real-valued functions $h_j(t), \hbar_j(t), t\in\mathbb{Z}$  such that
\begin{align*}
CA^{t-\tau}B=\sum\limits_{j=1}^{n}h_j(t)\hbar_j(\tau),\quad t,\tau\in\mathbb{Z},
\end{align*}
\end{Lemma}
\noindent \textit{Proof}:
The proof can be found in that of \cite[Theorem 1]{CA21}. 

First, the non-causal impulse response of the nominal model \eqref{eq:si_nominal_model} takes the form of \eqref{eq:noncausal_g0}.  
Then we have 
\begin{align*}
&g_0(t-\tau)= \left\{
\begin{array}{lll}
&C_c  A^{t-\tau-1}_cB_c,& \quad t-\tau\geq 1\\
&D-C_a  A^{-1}_aB_a, &\quad t-\tau= 0\\
&-C_a  A^{t-\tau-1}_aB_a, &\quad t-\tau\leq -1
\end{array}\right. 
\end{align*} 
where both $A_c$ and $A_a$ are non-singular. By Lemma \ref{le:se}, there exist four real-valued functions $h_{c,j}(t)$, $\hbar_{c,j}(t)$, $h_{a,j}(t)$, $\hbar_{a,j}(t)$, $t\in \mathbb{Z}$ such that 
\begin{align*}
&C_c  A^{t-\tau-1}_cB_c=\sum\limits_{j=1}^{n_{d,c}}h_{c,j}(t)\hbar_{c,j}(\tau),\quad  t, \tau\in \mathbb{Z},\\
&-C_a  A^{t-\tau-1}_aB_a=\sum\limits_{j=1}^{n_{d,a}}h_{a,j}(t)\hbar_{a,j}(\tau), \quad  t, \tau\in \mathbb{Z},
\end{align*}
where $n_{d,c}, n_{d,a} \in \mathbb{N}$  are the dimension of $A_c$ and $A_a$, respectively. 
Referring to Definition \ref{de:semise}, we first write down the generator functions of \eqref{eq:si_expre}, denoted by, $\mu_j(t), \nu_j(s)$, $t,s\in \mathbb{Z}$, $j=1,\cdots, n_{d,c}+ n_{d,a}$. For $j=1,\cdots, n_{d,c}$, we have 
\begin{align*}
&
\begin{array}{lll}
&\mu_j(t) = h_{c,j}(t),\\
&\nu_j(s) = M_{1,j}(s)+ M_{2,j}(s)+M_{3,j}(s),\end{array}
\\
&M_{1,j}(s)= \sum\limits_{j'=1}^{n_{d,c}} h_{c,j'}(s)  \sum\limits_{i=-\infty}^{s-1} b^2(i)\hbar_{c,j}(i)\hbar_{c,j'}(i),\\
&M_{2,j}(s)=\hbar_{c,j}(s)b^2(s)(D-C_aA_a^{-1}B_a),\\
&M_{3,j}(s)\!=\! \left\{
\begin{array}{lll}
& \sum\limits_{j'=1}^{n_{d,a}}h_{a,j'}(s) \sum\limits_{i=s+1}^{0}b^2(i)\hbar_{c,j}(i)\hbar_{a,j'}(i),& \text{ if }s\leq -1\\
&0,& \text{ if }s=0\\
&-\sum\limits_{j'=1}^{n_{d,a}}h_{a,j'}(s)\sum\limits_{i=1}^{s} b^2(i)\hbar_{c,j}(i)\hbar_{a,j'}(i),&\text{ if }s\geq 1
\end{array}\right.
\end{align*} and for $j=1,\cdots, n_{d,a}$, we have 
\begin{align*}
&
\begin{array}{lll}
&\mu_{n_{d,c}+j}(t) = N_{1,j}(t)+N_{2,j}(t)+ N_{3,j}(t),\\
&\nu_{n_{d,c}+j}(s) = h_{a,j}(s),\end{array}
\\
& N_{1,j}(t)=\sum\limits_{j'=1}^{n_{d,a}} h_{a,j'}(t) \sum\limits^{\infty}_{i=t+1} b^2(i)\hbar_{a,j}(i)\hbar_{a,j'}(i),\\
&N_{2,j}(t)= \hbar_{a,j}(t)b^2(t)(D-C_aA_a^{-1}B_a),\\
&N_{3,j}(t)\!=\! \left\{
\begin{array}{lll}
&\sum\limits_{j'=1}^{n_{d,c}}h_{c,j'}(t) \sum\limits_{i=1}^{t-1}b^2(i)\hbar_{a,j}(i)\hbar_{c,j'}(i), &\text{ if } t\geq 2\\
&0, &\text{ if }t=1\nonumber\\
&-\sum\limits_{j'=1}^{n_{d,c}}h_{c,j'}(t)\sum\limits_{i=t}^{0} b^2(i)\hbar_{a,j}(i)\hbar_{c,j'}(i), &\text{ if }t\leq 0
\end{array}\right.
\end{align*}

To prove Theorem \ref{th:se}, we need to show $k^{\text{NCSI}}(t,s)=\sum_{j=1}^{n_{d,c}+n_{d,a}}\mu_j(t)\nu_j(s)$, for $t>s$.
From  \eqref{eq:si_expre}, we have
\begin{align*}
&k^{\text{NCSI}}(t,s)= \sum\limits_{i=-\infty}^{s-1} b^2(i)g_0(t-i)g_0(s-i)\\
&+ \sum\limits_{i=s}^{t} b^2(i)g_0(t-i)g_0(s-i)+\sum\limits_{i={t+1}}^{\infty} b^2(i)g_0(t-i)g_0(s-i).\\
\end{align*} 
where 
\begin{subequations}
\begin{align*}
&\sum\limits_{i=-\infty}^{s-1} b^2(i)g_0(t-i)g_0(s-i) \\
&=\sum\limits_{j=1}^{n_{d,c}} h_{c,j}(t)\left(\sum\limits_{j'=1}^{n_{d,c}} h_{c,j'}(s)  \sum\limits_{i=-\infty}^{s-1} b^2(i)\hbar_{c,j}(i)\hbar_{c,j'}(i)\right)
\\&=\sum\limits_{j=1}^{n_{d,c}}\mu_j(t)M_{1,j}(s),
\end{align*}
\begin{align*}
&\sum\limits_{i=t+1}^{\infty} b^2(i)g_0(t-i)g_0(s-i) \\
&=\sum\limits_{j=1}^{n_{d,a}} h_{a,j}(s)\left(\sum\limits_{j'=1}^{n_{d,a}} h_{a,j'}(t) \sum\limits^{\infty}_{i=t+1} b^2(i)\hbar_{a,j}(i)\hbar_{a,j'}(i)\right)\\
&= \sum\limits_{j=1}^{n_{d,a}}N_{1,j}(t)\nu_{n_{d,c}+j}(s),
\end{align*}
\begin{align*}
&\sum\limits_{i=s}^{t} b^2(i)g_0(t-i)g_0(s-i)=b^2(s)g_0(t-s)g_0(0)\\
&\quad +\sum\limits_{i=s+1}^{t-1} b^2(i)g_0(t-i)g_0(s-i)+ b^2(t)g_0(0)g_0(s-t)\\
&=\sum\limits_{j=1}^{n_{d,c}}\mu_j(t)M_{2,j}(s)+\sum\limits_{j=1}^{n_{d,c}}\mu_j(t)M_{3,j}(s) \\
&\quad + \sum\limits_{j=1}^{n_{d,a}}N_{3,j}(t)\nu_{n_{d,c}+j}(s)+ \sum\limits_{j=1}^{n_{d,a}}N_{2,j}(t)\nu_{n_{d,c}+j}(s).
\end{align*}
\end{subequations}
Then note that, if any generator functions $\mu_j(t), \nu_j(s)$, $j=1,\cdots, n_{d,c}+ n_{d,a}$ vanish, the semiseparable rank will be smaller than $n_{d,c}+n_{d,a}$.
This completes the proof of Theorem \ref{th:se}.

To prove Corollary \ref{co:se}, we need to show $k^{\text{NCSI}}(t,t)=\sum_{j=1}^{n_{d,c}+n_{d,a}}\mu_j(t)\nu_j(t)$.
On one hand, we have from  \eqref{eq:si_expre}
\begin{align}
&k^{\text{NCSI}}(t,t)= \sum\limits_{i=-\infty}^{t-1} b^2(i)g_0(t-i)g_0(t-i)\nonumber\\
&+  b^2(t)g_0(0)g_0(0)+\sum\limits_{i={t+1}}^{\infty} b^2(i)g_0(t-i)g_0(t-i).\nonumber\\
&= \sum\limits_{j=1}^{n_{d,c}}\mu_j(t)M_{1,j}(t)+b^2(t)(D-C_aA^{-1}_aB_a)^2\nonumber\\
&+ \sum\limits_{j=1}^{n_{d,a}}N_{1,j}(t)\nu_{j+n_{d,c}}(t)\label{ap:h2tt_1}
\end{align}
On the other hand, we have 
\begin{align}
&\sum\limits_{j=1}^{n_{d,c}+n_{d,a}}\mu_j(t)\nu_j(t) =\sum\limits_{j=1}^{n_{d,c}}\mu_j(t)M_{1,j}(t)+ \sum\limits_{j=1}^{n_{d,c}}\mu_j(t)M_{2,j}(t)\nonumber\\
&+\sum\limits_{j=1}^{n_{d,c}}\mu_j(t)M_{3,j}(t) + \sum\limits_{j=1}^{n_{d,a}}N_{3,j}(t)\nu_{j+n_{d,c}}(t)\nonumber\\
&+ \sum\limits_{j=1}^{n_{d,a}}N_{2,j}(t)\nu_{j+n_{d,c}}(t)+ \sum\limits_{j=1}^{n_{d,a}}N_{1,j}(t)\nu_{j+n_{d,c}}(t)\nonumber\\
&=\sum\limits_{j=1}^{n_{d,c}}\mu_j(t)M_{1,j}(t)+ b^2(t)(C_cA_c^{-1}B_c)(D-C_aA_a^{-1}B_a)\nonumber\\
&-b^2(t)(C_cA_c^{-1}B_c)(-C_aA_a^{-1}B_a)\nonumber\\
&+ b^2(t)(-C_aA_a^{-1}B_a)(D-C_aA_a^{-1}B_a)\nonumber\\&+\sum\limits_{j=1}^{n_{d,a}}N_{1,j}(t)\nu_{j+n_{d,c}}(t) \label{ap:h2tt_2}
\end{align}
Clearly, \eqref{ap:h2tt_1} and \eqref{ap:h2tt_2} are not equal in general unless  
\begin{align*}
D(D-C_cA_c^{-1}B_c-C_aA_a^{-1}B_a)=0.
\end{align*}
This completes the proof of Corollary \ref{co:se}. 



\subsection{Closed Form of the NCSI-FO Kernel \eqref{eq:kernel_si_9}}\label{ap:ncsifo}
To avoid the abuse of notation, we let $k(t,s;\eta)$ represent $k^{\text{NCSI-FO}}(t,s;\eta)$ in this section. Note that the kernel is symmetric, so we only consider $k(t,s;\eta)$ with $t\geq s$. It follows from \eqref{eq:si_expre} that 
\begin{align*}
&k(t,s;\eta)=k_{c}(t,s;\eta)+ k_{\delta}(t,s;\eta)+ k_{a}(t,s;\eta),\\
&k_{c}(t,s;\eta)=\sum\limits_{k=1}^{\infty}\sigma^2_c \lambda_c^{k}g_0(t-k;\eta)g_0(s-k;\eta),\\
&k_\delta(t,s;\eta)=\sigma^2_0 g_0(t;\eta)g_0(s;\eta),\\
&k_{a}(t,s;\eta)=\sum\limits_{k=-\infty}^{-1}\sigma^2_a \lambda_a^{-k}g_0(t-k;\eta)g_0(s-k;\eta).
\end{align*}
Substituting \eqref{eq:ncsi_fo1} into  $g_0(\cdot)$, we obtain the closed form as follows.
\begin{align*}
&k_\delta(t,s;\eta)=\left\{
\begin{array}{lll}
 \sigma_0^2c_c^2a_c^{t+s},& t\geq s>0\\
 \sigma_0^2c_cc_0a_c^{t}, &t>s=0\\
 \sigma_0^2c_cc_aa_c^{t}a_a^{-s},& t>0>s\\
 \sigma_0^2c_0c_aa_a^{-s}, &t=0>s\\
 \sigma_0^2c_a^2a_a^{-t-s},& 0>t\geq s\\ 
\sigma_0^2c_0^2, &t=s=0
\end{array}\right. ,
\end{align*}
\begin{align*}
&k_c(t,s;\eta)=\left\{
\begin{array}{lll}
&\sum\limits_{i=1}^5 K_i(t,s), & t> s>1\\
&\sum\limits_{i=1}^4 K_i(t,s), & t>s=1\\
&\sum\limits_{i=1}^3 K_i(t,s), & t>1>s\\
&\sum\limits_{i=1}^2 K_i(t,s), &t=1>s\\
&\sum\limits_{i=1}^1 K_i(t,s), & 1>t> s\\
&K_1(t,t)+K_5(t,t)+K_6(t,t), &t=s>1\\
&K_1(t,t)+K_6(t,t), &t=s=1\\
&K_1(t,t), &1>t=s\\
\end{array}\right. ,
\end{align*}
with \begin{align*}
&K_1(t,s)=\sigma_c^2c_a^2 a_a^{-t-s}  \frac{(\lambda_ca_a^2)^{\max\{1,t+1\}}}{1-\lambda_ca_a^2},\\
&K_2(t,s)=  \sigma_c^2c_0c_a a_a^{t-s} \lambda_c^{t}  ,  \\
&K_3(t,s)= \sigma_c^2c_cc_a a_c^{t}a_a^{-s} \frac{(\lambda_ca_aa_c^{-1})^{\max\{s+1,1\}}-(\lambda_ca_aa_c^{-1})^t}{1-\lambda_ca_aa_c^{-1}}  ,\\
&K_4(t,s)= \sigma_c^2c_0c_c a_c^{t-s} \lambda_c^{s} ,\\
&K_5(t,s)= \sigma_c^2c_c^2a_c^{t+s}\frac{\lambda_ca_c^{-2} - (\lambda_ca_c^{-2})^s}{1-\lambda_ca_c^{-2}} , \\
&K_6(t,t)= \sigma_c^2 c_0^2\lambda_c^t .
\end{align*}
Note that $k_{a}(t,s;\eta)$ can be obtained in the same way as $k_{c}(t,s;\eta)$ because 
\begin{align*}
&k_a(t,s;\eta)=k_{c}(-t,-s;\eta^*), \\
&\eta = [a_c, a_a,c_c, c_0,c_a, \lambda_c,\lambda_a,\sigma_c, \sigma_a, \sigma_0],\\
&\eta^*= [a_a, a_c,c_a, c_0,c_c, \lambda_a,\lambda_c,\sigma_a, \sigma_c, \sigma_0].
\end{align*}

\begin{Remark}\label{re:sigma_0isone}
It is shown that $k_{c}(t,s)$, $k_{\delta}(t,s)$ and $k_{a}(t,s)$ all have the product $\sigma^2_{(\cdot)}c_{(\cdot)}c'_{(\cdot)}$ with $\sigma_{(\cdot)}\in \{\sigma_c,\sigma_a, \sigma_0$\} and $c_{(\cdot)},c'_{(\cdot)}\in \{c_c,c_a, c_0$\}. It follows that
\begin{align*}
&k^{\text{NCSI-FO}}(t,s;\eta_1)= k^{\text{NCSI-FO}}(t,s;\eta_2),\\
&\eta_1= [a_c, a_a,c_c, c_0,c_a, \lambda_c,\lambda_a,\sigma_c, \sigma_a, \sigma_0],\\
&\eta_2= [a_c, a_a,c_c\!\cdot\! \sigma_0, c_0\!\cdot\!\sigma_0,c_a\!\cdot\!\sigma_0, \lambda_c,\lambda_a,\frac{\sigma_c}{\sigma_0}, \frac{\sigma_a}{\sigma_0}, 1],
\end{align*}
This leads to the identifiability issue in the hyper-parameter estimation. Thus, $\sigma_0$ is always fixed to be 1 and not treated as the hyper-parameter. 
\end{Remark}

\end{document}